\documentclass[sigconf]{acmart}

\usepackage{appendix}
\usepackage{enumitem}
\usepackage[ruled,vlined]{algorithm2e}
\usepackage{subfigure}
\usepackage{multirow}
\usepackage{balance}
\usepackage[normalem]{ulem}

\usepackage{amsfonts,amssymb}
\usepackage{amsmath,bm}
\usepackage{threeparttable}
\newtheorem{myDef}{Definition}
\usepackage{color}
\usepackage{soul}

\AtBeginDocument{%
  \providecommand\BibTeX{{%
    \normalfont B\kern-0.5em{\scshape i\kern-0.25em b}\kern-0.8em\TeX}}}

\copyrightyear{2022} \acmYear{2022} \setcopyright{acmcopyright}\acmConference[KDD '22]{Proceedings of the 28th ACM SIGKDD Conference on Knowledge Discovery and Data Mining}{August 14--18, 2022}{Washington, DC, USA} \acmBooktitle{Proceedings of the 28th ACM SIGKDD Conference on Knowledge Discovery and Data Mining (KDD '22), August 14--18, 2022, Washington, DC, USA} \acmPrice{15.00} \acmDOI{10.1145/3534678.3539331} \acmISBN{978-1-4503-9385-0/22/08}

\acmConference[KDD '22]{The 28th ACM SIGKDD Conference on Knowledge Discovery and Data Mining}{August 14--18, 2022}{Washington, DC, USA}
\acmPrice{15.00}
\acmISBN{978-1-4503-XXXX-X/18/06}

\acmSubmissionID{1085}

\begin{document}

\newcommand{\MBLUE}[1]{\textcolor{blue}{#1}}

\newcommand{\MARK}[1]{#1}

\newcommand{\ClusterEAplain}[1]{ClusterEA#1}
\newcommand{\ClusterEA}[1]{\textsf{ClusterEA#1}}
\newcommand{\Sampling}[0]{ClusterSampler}
\newcommand{\Merging}[0]{SparseFusion}
\newcommand{\KMeansFullName}[0]{Cross-KG Mapping-based ClusterSampler}
\newcommand{\MetisFullName}[0]{Intra-KG Structure-based ClusterSampler}
\newcommand{\KMeans}[0]{CMCS}
\newcommand{\MetisGCN}[0]{ISCS}
\newcommand{\SparseCSLS}[0]{Sp-CSLS}

\newcommand{\HitNFull}[0]{Hits@$N$}
\newcommand{\HitN}[0]{H@$N$}
\newcommand{\HitOne}[0]{H@$1$}
\newcommand{\HitTen}[0]{H@$10$}
\newcommand{\MRR}[0]{MRR}

\title{\ClusterEA{}: Scalable Entity Alignment with Stochastic Training and Normalized Mini-batch Similarities}

\author{Yunjun Gao}
\authornote{Both authors contributed equally to this research.}
\affiliation{%
  \institution{Zhejiang University}
  \city{Hangzhou}
  \country{China}
}
\email{gaoyj@zju.edu.cn}


\author{Xiaoze Liu}
\authornotemark[1]
\affiliation{%
  \institution{Zhejiang University}
  \city{Hangzhou}
  \country{China}}
\email{xiaoze@zju.edu.cn}

\author{Junyang Wu}
\affiliation{%
  \institution{Zhejiang University}
  \city{Hangzhou}
  \country{China}}
\email{wujunyang@zju.edu.cn}

\author{Tianyi Li}
\affiliation{%
  \institution{Aalborg University}
  \city{Aalborg}
  \country{Denmark}}
\email{tianyi@cs.aau.dk}

\author{Pengfei Wang}
\affiliation{%
  \institution{Zhejiang University}
  \city{Ningbo}
  \country{China}}
\email{wangpf@zju.edu.cn}

\author{Lu Chen}
\affiliation{%
  \institution{Zhejiang University}
  \city{Hangzhou}
  \country{China}}
\email{luchen@zju.edu.cn}

\renewcommand{\shortauthors}{Gao and Liu et al.}


\begin{abstract}
Entity alignment (EA) aims at finding equivalent entities in different knowledge graphs (KGs). Embedding-based approaches have dominated the EA task in recent years.
Those methods face problems that come from the geometric properties of embedding vectors, including hubness and isolation. To solve these geometric problems, many normalization approaches have been adopted for EA.
However, the increasing scale of KGs renders it hard for EA models to adopt the normalization processes, thus limiting their usage in real-world applications. To tackle this challenge, we present \ClusterEA{}, a general framework that is capable of scaling up EA models and enhancing their results by leveraging normalization methods on mini-batches with a high entity equivalent rate.
\ClusterEA{} contains three components to align entities between large-scale KGs, including stochastic training, \Sampling{}, and \Merging{}. 
It first trains a large-scale Siamese GNN for EA in a stochastic fashion to produce entity embeddings. Based on the embeddings, a novel \Sampling{} strategy is proposed for sampling highly overlapped mini-batches. Finally, \ClusterEA{} incorporates \Merging{}, which normalizes local and global similarity and then fuses all similarity matrices to obtain the final similarity matrix.
Extensive experiments with real-life datasets on EA benchmarks offer insight into the proposed framework, and suggest that it is capable of outperforming the state-of-the-art scalable EA framework by up to $8$ times in terms of $Hits@1$.
\end{abstract}



\maketitle

\section{Introduction}
\label{sec:intro}
Knowledge graphs (KGs) represent collections of relations between real-world objects, which facilitate many downstream applications, such as semantic search~\cite{XiongPC17} and recommendation systems~\cite{ZhangYLXM2016}.
Although various KGs have been constructed in recent years, they are still highly incomplete. To be more specific, KGs built from different data sources hold unique information individually while having overlapped entities. This motivates us to integrate the knowledge of different KGs with the overlapped entities to complete KGs.
Entity alignment (EA)~\cite{MTransE17}, a fundamental strategy for knowledge integration, has been widely studied. EA aims to align entities from different KGs that refer to the same real-world objects, and thus, it facilitates the completion of KGs.

Embedding-based EA has been proposed~\cite{MTransE17}, and has been witnessed rapid development in recent years~\cite{KECG19, RREA20, AliNet20, HyperKA20, DualAMN21} thanks to the use of Graph Neural Networks (GNNs)~\cite{GCN17, GAT18, GraphSAGE17}.
They assume that the neighbors of two equivalent entities in KGs are also equivalent~\cite{AttrGNN20}. Based on this, they align entities by applying representation learning to KGs. We summarize the process of Embedding-based EA as the following three steps: (i) taking two input KGs and collecting \emph{seed alignment} as training data; (ii) training an EA model with the isomorphic graph structure of two KGs to encode entities into embedding vectors; and (iii) aligning the equivalent entities between the two input KGs based on a specific similarity measurement (e.g., cosine similarity) of their corresponding embeddings.

The size of real-world KGs is much larger than that of conventional datasets used in evaluating EA tasks.
For instance, a real-world KG YAGO3 includes 17 million entities~\cite{LargeEA22}. Thus, EA methods should be scaled up to massive data in order to adapt to real-world applications. However, a recent proposal~\cite{LargeEA22}
lost too much graph structure information, trading the quality of results for scalability.
Worse still, as the input KGs become larger, using greedy search~\cite{OpenEA2020VLDB} to find corresponding entities \MARK{from one KG to another} with top-1-nearest neighbor becomes more challenging.
%
Specifically, the geometric properties of high-dimensional embedding vector spaces lead to problems for embedding-based EA, namely, \emph{geometric problems}, mainly including the hubness and isolation problems~\cite{OpenEA2020VLDB}.
The hubness problem indicates that some points (known as hubs) frequently appear as the top-1 nearest neighbors of many other points in the vector space.
The isolation problem implies that some outliers would be isolated from any point clusters. As the scale of input KGs grows, these problems become even more severe due to the increasing number of candidates of nearest neighbor search.
EA is generally assumed to follow 1-to-1 mapping~\cite{MTransE17}.
Many existing methods have been proposed to solve the geometric problems by making the similarity matrix better satisfy this assumption, i.e., \emph{normalization} methods.
A list of widely used normalization methods for EA includes (i) \emph{Cross-domain Similarity Local Scaling (CSLS)}~\cite{CSLS} that is adopted from word translation~\cite{TransEdge19,EVA20, RREA20, EASY21}; (ii) \emph{Gale-Shapley algorithm}~\cite{GaleShapley} that treats EA as stable matching~\cite{CEAFF20,CEAFF21}; (iii) \emph{Hungarian algorithm}~\cite{Hungarian1955} and (iv) \emph{Sinkhorn iteration}~\cite{Sinkhorn13} that both transform EA as the assignment problem~\cite{DGMC20, SEU21, EASY21}.
Specifically, Sinkhorn iteration~\cite{Sinkhorn13} could significantly improve the accuracy of matching entities with embeddings~\cite{EASY21, SEU21}. Moreover, Sinkhorn iteration suits the normalization approach of EA best as it can be easily parallelized on the GPU. 
Nonetheless, the existing normalization approaches~\cite{CSLS,Sinkhorn13, GaleShapley, Hungarian1955} are at least of quadratic complexity. This prohibits them from being applied to large-scale data, thus limiting their real-world applications.

In this work, we aim to scale up the normalization process of the similarity matrix to achieve higher EA performance.
We adopt a standard machine learning technique, sampling mini-batches, to perform EA in linear time.
Specifically, we first train a GNN model to obtain the global embeddings of two KGs.
Then, we generate mini-batches for two KGs by placing entities that could find their equivalent together.
\MARK{Next, we calculate and normalize a local similarity matrix between two sets of entities selected for each mini-batch by Sinkhorn iteration.}
Finally, we merge the local similarities into a unified and sparse similarity matrix.
With this strategy, the final similarity matrix is normalized with Sinkhorn iteration, thus achieving higher accuracy, and the time and space complexities can also be significantly reduced.
However, its materialization is non-trivial due to the two major challenges:
\begin{itemize}[topsep=0pt,itemsep=0pt,parsep=0pt,partopsep=0pt,leftmargin=*]
  \item \textit{How to sample mini-batches with high entity equivalence rate to ensure 1-to-1 mapping?} Splitting mini-batches on two KGs is quite different from conventional tasks. To transfer EA within mini-bathes into the assignment problem, we should place possibly equivalent entities into the same batch to meet the 1-to-1 mapping assumption. Nevertheless, the mapping is only partially known (as the training set) for the EA task, making it hard for two entities to be aligned and hence to be placed in the same batch. Intuitively, randomly splitting two KGs will make most mini-batches fail to correspond. An existing study~\cite{LargeEA22} proposes a rule-based method to split batches with a higher rate of entity equivalence. Nonetheless, its results are still not enough to satisfy the 1-to-1 mapping assumption.
  \item \textit{How to fuse the mini-batch similarity matrices to ensure high accuracy?} The similarity matrix of each batch focuses only on its local information. Thus, the results generally deviate from those obtained without sampling, even if the batch sampler is good enough. This motivates us to study two problems. First, how to combine the similarities of the mini-batches to get a better global optimal match, after getting the mini-batch. Second, how to improve the accuracy of results obtained by batch samplers.
\end{itemize}

\begin{figure}[t]
\centering 
\hspace{-4.8mm}\includegraphics[width=3.55in]{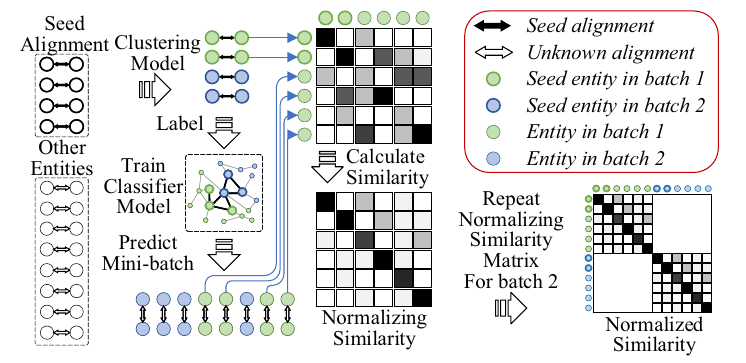} 
\caption{An example of normalizing mini-batch similarity matrix with \Sampling{} strategy}
\vspace{-5mm}
\label{fig:example}
\vspace{-3mm}
\end{figure}

To tackle the aforementioned two challenges, we present a general EA framework, \ClusterEA{}, that can be applied to arbitrary GNN-based EA models while can be able to achieve high accuracy and scalability.
To scale up GNN training over large-scale KGs, \ClusterEA{} first utilizes neighborhood sampling~\cite{GraphSAGE17} to train a large-scale GNN for EA in a stochastic fashion.
Then, with the embedding obtained from the GNN model, \ClusterEA{} proposes a novel \emph{strategy} for generating high-quality mini-batches in EA, called \Sampling{}. As depicted in Figure~\ref{fig:example}, \Sampling{} first labels the training pairs into different batches with a \emph{clustering} method. Then, it performs supervised \emph{classification} using previously labeled training pairs to generate mini-batches for all the entities. By changing the \emph{clustering} and \emph{classification} models, the strategy can sample mini-batches by capturing multiple aspects of the graph structure information. We propose two sampling methods within \Sampling{}, including
(i)  \emph{\MetisFullName{} (\MetisGCN{})} that retains the intra-KG graph structure information such as the neighborhood of nodes; and
(ii) \emph{\KMeansFullName{} (\KMeans{})} which retains the inter-KG matching information provided by the learned embedding.
These methods sample two KGs into multiple mini-batches with a high entity equivalent rate to satisfy the 1-to-1 mapping assumption better.
Finally, \ClusterEA{} uses our proposed \Merging{}, which fuses the similarity matrices calculated separately using Sinkhorn iteration with normalized global similarity. The \Merging{} produces a matrix with high sparsity while keeping as much valuable information as possible.
Our contributions are summarized as follows:

\begin{itemize}[topsep=0pt,itemsep=0pt,parsep=0pt,partopsep=0pt,leftmargin=*]
    \item{\emph{Scalable EA framework.}}
    We develop \ClusterEA{}\footnote{https://github.com/joker-xii/ClusterEA}, a scalable framework for EA, which reduces the complexity of normalizing similarity matrices with mini-batch. To the best of our knowledge, this is the first framework that utilizes stochastic GNN training for large-scale EA. Any GNN model can be easily integrated into \ClusterEA{} to deal with large-scale EA (Section~\ref{sec:framework}) with better scalability and higher accuracy.
    \item{\emph{Fast and accurate batch samplers.}}
    We present \Sampling{}, a novel \emph{strategy} that samples batches by learning the latent information from embeddings of the EA model. In the strategy, we implement two samplers to capture different aspects of KG information, including
    (i) \MetisGCN{} that aims at retaining the intra-KG graph structure information, and
    (ii) \KMeans{} that aims at retaining the inter-KG alignment information. Unlike the previous rule-based method, the two samplers are learning-based, and thus can be parallelized and produce high-quality mini-batches.
    \item{\emph{Fused local and global similarity matrix.}} We propose \Merging{} for normalizing and fusing not only local similarity matrices but also global similarity matrices into one unified sparse matrix, which enhances the expressive power of EA models.
    \item{\emph{Extensive experiments.}} We conduct comprehensive experimental evaluation on EA tasks compared against state-of-the-art approaches over the existing EA benchmarks. Considerable experimental results demonstrate that  \ClusterEA{} successfully achieves satisfactory accuracy on both conventional datasets and large-scale datasets (Section~\ref{sec:exp}).
\end{itemize}

\vspace*{-2mm}
\section{Related Work}
\label{sec:related_work}

Most existing EA proposals find equivalent entities by measuring the similarity between the embeddings of entities. Structures of KGs are the basis for the embedding-based EA methods.
Representative EA approaches that rely purely on KGs' structures can be divided into two categories, namely, \emph{KGE}-\emph{based EA}~\cite{MTransE17, IPTransE17, BootEA18, TransEdge19} and \emph{GNN-based EA}~\cite{GCN-Align18, KECG19, MRAEA20, AliNet20, HyperKA20, MuGNN19}.
The former incorporates the KG embedding models (e.g., TransE~\cite{TransE13}) to learn entity embeddings. The latter
learns the entity embeddings using GNNs~\cite{GCN17}, which aggregates the neighbors' information of entities.

In recent years, GNN-based models have demonstrated their outstanding performance~\cite{DualAMN21}. This is contributed by the strong modeling capability on the non-Euclidean structure of GNNs with anisotropic attention mechanism~\cite{GAT18}. Nonetheless, they suffer from poor scalability~\cite{LargeEA22} due to the difficulty in sampling mini-batches with neighborhood information on KGs. LargeEA~\cite{LargeEA22}, the first study focusing on the scalability of EA, proposes to train GNN models on relatively small batches of two KGs independently.
The small batches are generated with a rule-based partition strategy called METIS-CPS.
However, massive information of both graph structures and seed alignment is lost during the process, resulting in poor structure-based accuracy. In contrast, \ClusterEA{} utilizes neighborhood sampling~\cite{GraphSAGE17}. Specifically, it trains one unified GNN model on the two KGs, during which the loss of structure information is neglectable.
Moreover, \ClusterEA{} creates mini-batches using multiple aspects of graph information, resulting in better entity equivalent rate than METIS-CPS.

In addition to structure information, many existing proposals facilitate the EA performance by employing \emph{side information} of KGs,  including
\emph{entity names} \cite{JAPE17,MultiKE19,DGMC20,DegreeAware20,BERT-INT20, AttrGNN20, EASY21, SEU21, LargeEA22, DSE1,DSE2}, \emph{descriptions} \cite{MultiKE19, BERT-INT20}, \emph{images} \cite{EVA20}, and \emph{attributes} \cite{JAPE17,GCN-Align18,MultiKE19,COTSAE20,BERT-INT20,AttrGNN20,EPEA20}. Such proposals are able to mitigate the geometric problems~\cite{OpenEA2020VLDB}.
Nonetheless, the models using side information mainly have two main limitations. \MARK{First, side information may not be available due to privacy concerns}, especially for
industrial applications~\cite{DualAMN21,RREA20, MRAEA20}.
Second, \MARK{models that incorporating machine translation or pre-aligned word embeddings may be overestimated due to the name bias issue}~\cite{JEANS20, AttrGNN20, EVA20, NoMatch21}.
\MARK{Thus, compared with the models employing side information, the structure-only methods are more general and not affected by bias of benchmarks}.
To this end, we do not incorporate side information in \ClusterEA{}.

In order to solve the geometric problems of embedding-based EA approaches, CSLS~\cite{CSLS} have been widely adopted, which normalizes the similarity matrix in recent studies~\cite{TransEdge19,EVA20, RREA20, EASY21}.
However, \MARK{CSLS does not perform full normalization of the similarity matrix.
As a result, its improvement over greedy search of top-1-nearest neighbor is limited.}
CEA~\cite{CEAFF20, CEAFF21} adopts the Gale-Shapley algorithm~\cite{GaleShapley} to find the stable matching between entities of two KGs, which produces higher-quality results than CSLS.
Nevertheless, the Gale-Shapley algorithm is hard to be parallelized, and hence, it is almost infeasible to perform large scale EA. Recent studies have transformed EA into the assignment problem~\cite{EASY21, SEU21}. They adopt Hungarian algorithm~\cite{Hungarian1955} or Sinkhorn iteration~\cite{Sinkhorn13} to normalize the similarity matrix. However, the computational cost of such algorithms is high, prohibiting them from being applied to large scale EA.
\ClusterEA{} also utilizes Sinkhorn iteration~\cite{Sinkhorn13} which performs full normalization on the similarity matrix with GPU acceleration. Moreover, \ClusterEA{} develops novel batch-sampling methods for adopting the normalization to large-scale datasets with the loss of information being minimized.

\section{Preliminaries}
\label{sec:define}
We proceed to introduce preliminary definitions. Based on these, we formalize the problem of entity alignment.

\vspace{-2mm}
\begin{myDef}
A \textbf{knowledge graph} (KG) can be denoted as $G = (E,R,T)$, where $E$ is the set of entities, $R$ is the set of relations, and $T=\{(h,r,t)~|~h,t \in E, r \in R\}$ is the set of triples, each of which represents an edge from the head entity $h$ to the tail entity $t$ with the relation $r$.
\end{myDef}

\vspace{-5mm}
\begin{myDef}
\label{sec:problem_statement}
\textbf{Entity alignment} (EA) \cite{OpenEA2020VLDB} aims to find the 1-to-1 mapping of entities $\phi$ from a source KG $G_s = (E_s,R_s,T_s)$ to a target KG $G_t = (E_t,R_t,T_t)$.
Formally, $\phi = \{(e_s, e_t) \in E_s \times E_t~|~e_s \equiv e_t\}$, where
$e_s \in E_s$, $e_t \in E_t$, and $\equiv$ is an equivalence relation between $e_s$ and $e_t$.
In most cases, a small set of equivalent entities $\phi^{\prime} \subset \phi$ \MARK{is known beforehand and} used as seed alignment.
\end{myDef}

\vspace{-5mm}
\begin{myDef}
\textbf{Embedding-based EA} aims to learn a set of embeddings for all entities $E_s$ and $E_t$, denoted as $\mathbf{f}\in \mathcal{R}^{(|E_s|+|E_t|)\times D}$, and then, it tries to maximize the similarity (e.g. cosine similarity) of entities that are equivalent in $\phi$, where $D$ is the size of embedding vectors.
\end{myDef}

\section{Our Framework}
\label{sec:framework}
In this section, we present our proposed framework \ClusterEA{}, a novel scalable EA framework. We start with the overall framework, followed by details on each component of our framework.

\subsection{Overall Framework of \ClusterEAplain{}}

\begin{figure*}[t]
\vspace*{-4mm}
\centering
\includegraphics[width=5.6in]{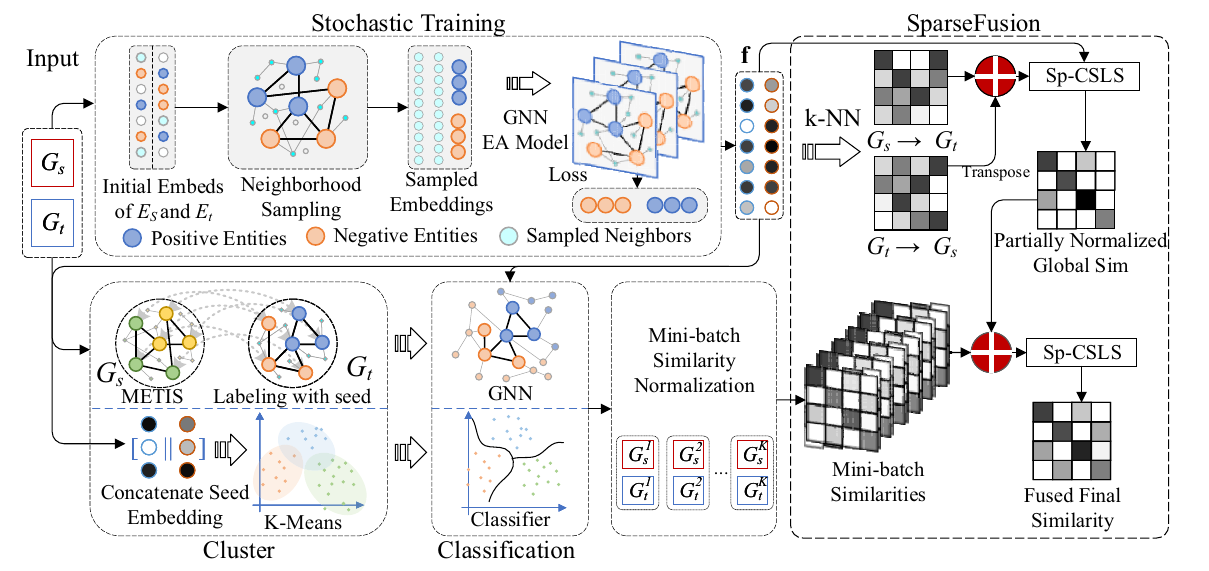}
\vspace*{-4mm}
\caption{The overall \ClusterEAplain{} framework}
\label{fig:framewrok}
\vspace*{-4mm}
\end{figure*}

As shown in Figure~\ref{fig:framewrok}, to scale up GNN training over large-scale KGs, \ClusterEA{} first utilizes neighborhood sampling~\cite{GraphSAGE17} to train a large-scale GNN for EA in a stochastic fashion.
Thereafter, \ClusterEA{} proposes a novel learning-based batch sampling strategy \Sampling{}. It uses the embedding vectors obtained from stochastic training. In the strategy, two batch samplers that learn multiple aspects of the input KGs are applied to the input KGs, including
(i) \MetisGCN{} that aims at retaining the intra-KG graph structure information such as the neighborhood of nodes, and
(ii) \KMeans{} that aims at retaining the inter-KG matching information provided by the learned embedding.
These methods sample two KGs into multiple mini-batches with a high entity equivalent rate that better satisfies the 1-to-1 mapping assumption.
Finally, \ClusterEA{} proposes \Merging{}, which fuses the normalized local similarity matrices with partially normalized global similarity. The \Merging{} produces the fused final matrix with high sparsity while keeping as much valuable information as possible.

\vspace{-3mm}
\subsection{Stochastic Training of GNNs for EA}
\label{sec:mini-batch-training}

GNN-based methods \cite{AttrGNN20,KECG19, EVA20, AliNet20, DualAMN21} have dominated the EA tasks with promising performances by propagating the information of seed alignments to their neighbors.
Inspired by this, we propose to incorporate GNN-based models into \ClusterEA{}.
\ClusterEA{} provides a general framework for training a Siamese GNN on both $G_s$ and $G_t$ that all GNN-based EA models follow~\cite{DualAMN21}. As a result, any existing GNN model can be used in \ClusterEA{} to produce the structural feature embeddings of entities.

To scale up the existing GNN-based EA models, \ClusterEA{} trains the models with neighborhood sampling. We sample mini-batches based on the seed alignment. Specifically, following the negative sampling process, we first randomly select a mini-batch of size $N_p$ containing source and target entities $\phi_s'$ and $\phi_t'$ in the seed alignment $\phi^{\prime}$ that are equivalent. Then, we randomly sample source and target entities size of $N_n$ with seed alignment in their whole entity sets that does not overlap with selected seed entities, denoted as $\theta_s = \{e_s |e_s \in E_s \cap e_s \not\in \phi_s'\}$ and $\theta_t = \{e_t |e_t \in E_t \cap e_t \not\in \phi_t'\}$. Finally, a mini-batch $B = \{B_s, B_t\} = \{(\phi_s' \cup \theta_s), (  \phi_t' \cup \theta_t) \}$ is generated waiting for the GNN model to produce its embeddings.

Generally, the GNN-based models train the entity's embedding by propagating the neighborhood information~\cite{RREA20, GCN17, GAT18}.
Formally, the embedding of an entity $v \in B$ in the $k_{th}$ layer of GNN $h_v^k$ is obtained by aggregating localized information via

\vspace{-3mm}
\begin{equation}
\begin{array}{c}
a_{v}^{(k)}=\operatorname{Aggregate}^{(k)}(\left\{h_{u}^{(k-1)} \mid u \in \mathcal{N}(v)\right\}) \\
h_{v}^{(k)}=\operatorname{Update}^{(k)}(a_{v}^{(k)}, h_{v}^{(k-1)})
\end{array}
\label{eq:message_passing}
\vspace{-1mm}
\end{equation}
where $h_v^0 \in \mathcal{R}^D$ is a learnable embedding vector initialized with Glorot initialization, and 
$\mathcal{N}(v)$ represents the set of neighboring entities around $v$. The model's final output on entity $e$ is denoted as $\mathbf{f}_e$. By applying neighborhood sampling, the size of neighborhood in each GNN layer of each entity is limited that no more than a fan out hyperparameter $F$, formally $|\mathcal{N}(v)|\leq F$. We place the graph information on CPU memory. When computing one layer of GNN in each batch, the neighbor of entities in current batch is sampled to form a block of graph, the graph information of this block will be transferred to GPU memory, along with the computation graph, making the final loss backward propagated with GPU.

To maximize the similarities of equivalent entities in each mini-batch, GNN-based EA models often use triplet loss along with negative sampling~\cite{GCN-Align18, KECG19, RREA20, MRAEA20}. In this paper, the \emph{Normalized Hard Sample Mining (NHSM)} loss~\cite{DualAMN21}, an extension for negative sampling that could significantly reduce training epochs, is adopted by us for training large-scale GNNs. We detail how we apply the NHSM loss in Appendix~\ref{app:nhsm}.

\noindent
\textbf{Discussions.}
Due to the neighborhood sampling process, our stochastic version of EA training will have a certain graph information loss, which is minimized with the randomness of sampling. Recent studies~\cite{ClusterGCN} have proposed to limit the sampling in small fixed sub-graphs for better training speed, which will further decrease the accuracy.
LargeEA~\cite{LargeEA22}, on the other hand, trains multiple GNNs on small batches generated with a rule-based method. Such an approach could fasten the training process. Nonetheless, much of the structure information is lost during partitioning, incurring poor performance. Although LargeEA can scale up the EA models to deal with large-scale datasets, it has too much trade-off on the accuracy, which is unreasonable.
\subsection{Learning-based Mini-batch Samplers}

After obtaining the KG embeddings, we aim to build batch samplers that utilize the features learned by EA models for generating mini-batches with high entity equivalent rates. To this end, we present the \Sampling{} strategy, which first \emph{clusters} the training set for obtaining batch labels, and then fits a \emph{classification} model with train labels to put all the entities into the right batch. The two models must satisfy two rules: (i) \emph{scalability} that both the classification and clustering method should be able to apply on large-scale data; and (ii) \emph{distinguishability} that the classifier must be able to distinguish entities into different labels the clustering method provides. If the \emph{scalability} rule is not satisfied, the model will crash due to limited computing resources. If the \emph{distinguishability} rule is not satisfied, the model cannot produce reasonable output. For example, if we randomly split the train set, there is no way for any model to classify the entities with such a label. Following the two rules, by changing different clustering and classification methods, we propose two batch samplers capturing different aspects of information in the two KGs. Each of them produces a set
containing $K$ batches, where $K$ is a hyperparameter. Intuitively, larger $K$ would result in smaller batches, making the normalization process consumes less memory. However, it also makes the \Sampling{} process more difficult. We detail the effect of different $K$ on the accuracy of \Sampling{} in Section~\ref{sec:exp_minibatchgeneration}.  To distinguish the batches from batches in Section~\ref{sec:mini-batch-training}, the batches are denoted as $\mathcal{B} = \{(\mathcal{B}_s, \mathcal{B}_t)\}$, where $\mathcal{B}_s \subset E_s$ and $\mathcal{B}_t \subset E_t$ are the sets of source and target entities respectively for each batch.

\noindent
\textbf{Mini-batch sampling with Intra-KG information.} In the \Sampling{} strategy, we first present  \MetisFullName{} (\MetisGCN{}) for sampling batches based on learning neighborhood information of KGs. Following the  \emph{distinguishability} rule, for retaining the neighborhood information, the clustering method should put nodes that are neighbors into the same batch. This brings us to minimize the edge cut. A previous study proposes METIS-CPS~\cite{LargeEA22}, which clusters two KGs with METIS~\cite{METIS98}, a classic algorithm to partition large graphs for minimizing the edge cut. METIS-CPS is designed to minimize both the edge cuts of two KGs and the decrease of entity equivalent rate. It first clusters source KG with METIS, and then clusters target KG with higher weights set for train nodes guiding the METIS algorithm. However, the METIS algorithm on the target KG is also a clustering algorithm, thus does not necessarily follow the guidance of train nodes. Following the \Sampling{} strategy, \MetisGCN{} also adopts METIS for \emph{clustering} source KG, but trains a GNN~\cite{GCN17} for \emph{classifying} target KG. The labels provided by METIS will keep nodes that neighbor together as much as possible, and thus can be learned with neighborhood propagation of GNNs. The training and inference on target KG could follow standard node classification task settings. For scalability consideration, we adopt the classic GCN as the classification model. We use the learned embeddings $\mathbf{f}$ of the EA model as the input feature and cross-entropy as the learning loss to train a two-layer GCN. Since GCN does not recognize different relation types in KGs, we adopt the computation of weights in the adjacency matrix from GCNAlign~\cite{GCN-Align18} to convert triples with different relation types into different edge weights. We detail the adjacency matrix construction in Appendix~\ref{app:adj-matrix}.

\noindent
\textbf{Mini-batch sampling with Inter-KG information.} Recall that the cross-KG mapping information is learned into two sets of embeddings $\mathbf{f}_s$ and $\mathbf{f}_t$, partitioning based on these embeddings could preserve the mapping information as much as possible. We propose \KMeansFullName{} (\KMeans{}) for partitioning directly based on the embedding vectors.
For obtaining labels of training sets, in \emph{clustering} process of \KMeans{}, we adopt the K-Means algorithm, a widely-used and scalable approach to cluster embeddings in high-dimensional vector spaces.
K-Means may lead to the entities unevenly distributed in different batches. To reduce this effect and to obtain more balanced mini-batches, we normalize the entity features with the standard score normalization. We concatenate the normalized embeddings of the training set into one unified set of embeddings. Then, we cluster the embeddings to obtain the labeled batch number for the training set $C'$. Formally, 
$C'_{e_s} = C'_{e_t} = \operatorname{k-Means}(\mathbf{f}_{n}(e_s,e_t), K), \; (e_s, e_t) \in \phi'$, $\mathbf{f}_{n}(e_s,e_t)= [\operatorname{z-score}(\mathbf{f}_{e_s})||\operatorname{z-score}(\mathbf{f}_{e_t})]$, 
where $\operatorname{z-score}(X)=  \frac{X - \mu(X)}{\sigma(X)}$ is the standard score normalization.

Next, we use the label to train two classifiers for both $E_s$ and $E_t$, and predict on all the embeddings. We describe how to obtain the batch number of each entity as follows: $C_{E_s} = \{ \operatorname{clf}(\mathbf{f}_{\phi'_s}, C'_{\phi'_s}, \mathbf{f}_{e_s}) \;|\; e_s \in E_s \}$ and $
    C_{E_t} = \{ \operatorname{clf}(\mathbf{f}_{\phi'_t}, C'_{\phi'_t}, \mathbf{f}_{e_t}) \;|\; e_t \in E_t \}$, 
where $\operatorname{clf}(\mathbf{f}_{train}, C_{train}, \mathbf{f})$ denotes the classification model. It trains based on the training set embedding $\mathbf{f}_{train}$ and label $C_{train}$, and then, it predicts the class for embedding $\mathbf{f}$. In this paper, we use XGBoost Classifier~\cite{XGBoost16}, a scalable classifier that has been a golden standard for various data science tasks. After classification, we can easily obtain the batches with the label $C$.

\subsection{Fusing Local and Global Similarities}

Since the \Sampling{} strategy utilizes different aspects of information to learn the mini-batches, the mini-batch similarity matrices generated may be biased by the corresponding batch sampler. For example, \KMeans{} only relies on the embeddings, tending to put entities with similar embeddings together. This information bias may have a negative effect on the final accuracy. To avoid such bias as much as possible, we propose \Merging{}. It first applies Sinkhorn iteration on mini-batch similarity matrices generated by multiple batch samplers. Then, \Merging{} sums all the similarity matrices of generated batches to obtain a fused local similarity matrix. Finally, it further fuses the local similarity matrix with a partially normalized global similarity based on a newly proposed sparse version of CSLS~\cite{CSLS}, namely, \SparseCSLS{}.

\noindent
\textbf{Local Similarity Matrix Normalization.}
\label{sec:normalize-local-sim}
Previous section describes how to sample the input KGs into multiple batches. For each batch generated from a batch-sampler $\mathcal{B}^i \in \mathcal{B} = \{\mathcal{B}^i_s, \mathcal{B}^i_t\}, i \in K$,  we assume that there exists 1-to-1 mapping between the source and target entities. We first obtain the local similarity matrix $\mathcal{M}^i\in \mathcal{R}^{|E_s|\times|E_t|}$ of current batch.
Formally,

\begin{equation}
   \mathcal{M}^i_{e_s, e_t}  =
    \begin{cases}
      \operatorname{sim}(e_s, e_t) & \text{if $e_s \in \mathcal{B}^i_s$ and $e_t \in \mathcal{B}^i_t$}\\
      0 & \text{otherwise}
    \end{cases}
\end{equation}
where $\operatorname{sim}(e_{s}, e_{t})= \mathbf{h}_{e_s} \cdot \mathbf{h}_{e_t}$ is the similarity of two entities obtained with the GNN output feature.

Then, we follow~\cite{Sinkhorn13} to implement the Sinkhorn iteration. We iteratively normalize the similarity matrix $K_s$ rounds in each batch, converting the similarity matrix into a doubly stochastic matrix.
The entities of mini-batches in one graph do not have overlap with each other, meaning that there will be no overlapped values in all the mini-batch similarities. Therefore, to obtain the locally normalized similarity for the whole dataset, we directly sum up all the mini-batch similarities, denoted as $\mathcal{M} = \sum_{i \in K} \operatorname{Sinkhorn} (\mathcal{M}^i, K_s) \in [0,1] ^{|E_s|\times|E_t|}$.

\noindent
\textbf{Fusing multi-aspect local similarities.}
To avoid the bias from one batch sampler, we calculate multiple similarity matrices as described above with different batch samplers. We obtain the cross-KG information-based similarity $\mathcal{M}_C$ with \KMeans{}, and intra-KG information-based similarity $\mathcal{M}_I$ with \MetisGCN{}.
Since the \MetisGCN{} process is unidirectional, indicating that this process on $G_s \rightarrow G_t$ produces different result with $G_t \rightarrow G_s$. According to this characteristic, we apply \MetisGCN{} on both direction, resulting into two matrices $\mathcal{M}_{I, G_s\rightarrow G_t}$ and $\mathcal{M}_{I, G_t\rightarrow G_s}$. Following~\cite{CEAFF20}, we sum up all the similarity matrices without setting any weight to obtain the final local similarity matrix. Formally, $ \mathcal{M}_{L} = \mathcal{M}_C + \mathcal{M}_{I, G_s\rightarrow G_t}+ \mathcal{M}_{I, G_t\rightarrow G_s}^T$.
This simple approach of fusing multi-view similarity matrices is proved to be useful in various previous studies~\cite{CEAFF20, CEAFF21, EASY21, LargeEA22}.

\noindent
\textbf{Normalize global similarity with \SparseCSLS{}.} To fuse the normalized local similarity matrix $\mathcal{M}_{L}$ with the global similarity, we first obtain the global similarity, and normalize it partially.
A widely-used normalization approach for solving geometric problems is to apply CSLS~\cite{CSLS} on the similarity matrix.
Formally, for two entities $e_s$ and $e_t$,
$\operatorname{CSLS}(e_s, e_t) = 2\operatorname{sim}(e_s, e_t) - r_S(e_t) - r_T(e_s)$, where $r_S$ and $r_T$ are the nearest neighborhood similarity mean, which can be obtained by k-NN search with $K_n$ as the neighborhood size.


However, CSLS also lack scalability, which motivates us to propose a sparse version of CSLS, i.e., \SparseCSLS{}. Recent studies~\cite{LargeEA22, NoMatch21} apply FAISS~\cite{JDH17} to compute K-nearest neighbours of $E_s$ on the embedding space $\mathbf{h}_t$. Similar to the dense version of CSLS, \SparseCSLS{} normalizes a sparse similarity matrix, resulting in a partially normalized similarity matrix. It first uses FAISS to calculate mean neighborhood similarities  $r_{T}( x_{s})$ and $r_{S}(y_{t})$. Then, for a sparse matrix $\mathcal{M}$, the \SparseCSLS{} only subtracts nonzero values of $2 \mathcal{M}$ with $r_{T}( x_{s})$ and $r_{S}(y_{t})$, the result is denoted as $\mathcal{M}'$. To keep the nonzero values to be useful, the final output is normalized with min-max normalization. Formally, $\operatorname{Sp-CSLS}(\mathcal{M}) =\frac{ \mathcal{M}' - \operatorname{min}(\mathcal{M}')}{ \operatorname{max}(\mathcal{M}') - \operatorname{min}(\mathcal{M}')}$.

To obtain the normalized global similarity. We first utilize FAISS to obtain an initial global similarity matrix by fusing k-NN similarity matrices on both $\mathbf{f}_s \rightarrow \mathbf{f}_t$ and $\mathbf{f}_t \rightarrow \mathbf{f}_s$ directions. Next, we normalize it with \SparseCSLS{}. Formally, $\mathcal{M}_{G} = \operatorname{Sp-CSLS}(\operatorname{k-NN}(\mathbf{h}_s, \mathbf{h}_t,K_r) + \operatorname{k-NN}(\mathbf{h}_t, \mathbf{h}_s,K_r)^T, K_n)$, where $\operatorname{k-NN}(\mathbf{X}, \mathbf{Y},K_r)$ returns the similarity matrix of $\mathbf{X} \rightarrow \mathbf{Y}$ remaining only top-$K_r$ values, and $K_n$ is the neighborhood size for CSLS. Note that the sparse similarity of two directions may have some values overlapped, this overlapped values will become twice the value before. However, it will hardly have negative impact on accuracy since entity pairs that both have high ranking on the another KG are more possible to be correctly aligned~\cite{MRAEA20}.

\noindent
\textbf{Fusing normalized similarities.}
To obtain the fused final similarity matrix, we first fuse the global and local similarities, and then apply the \SparseCSLS{} for further normalization on the final matrix. Formally, $\mathcal{M}_{F}=\operatorname{Sp-CSLS}( \mathcal{M}_L +\mathcal{M}_G, K_n)$.

\section{Experiments}
\label{sec:exp}

In this section, we report on extensive experiments aimed at evaluating the performance of  \ClusterEA{}.

\vspace{-2mm}
\subsection{Experimental Settings}\label{sec:exp_setting}

\begin{table*}[t]\small
\vspace*{-4mm}
\caption{Overall EA results on IDS15K and IDS100K
}\label{exp:overall_15K_100K}
\vspace*{-4mm}
\begin{threeparttable}
\setlength{\tabcolsep}{0.6mm}{
\begin{tabular}{l|ccccc|ccccc|ccccc|ccccc}
\toprule
\multicolumn{1}{c|}{\multirow{2}{*}{Methods}} & \multicolumn{5}{c|}{IDS15K$_{EN-FR}$} & \multicolumn{5}{c|}{IDS15K$_{EN-DE}$} & \multicolumn{5}{c|}{IDS100K$_{EN-FR}$} & \multicolumn{5}{c}{IDS100K$_{EN-DE}$} \\ \cline{2-21}
 &  H@1 &  H@10 & MRR & Time &  Mem. & H@1 &  H@10 & MRR & Time &  Mem. &  H@1 &  H@10 & MRR & Time  &  Mem. &  H@1 &  H@10 & MRR & Time  &  Mem.\\ \hline
GCNAlign   & 38.2 & 78.5 & 0.51 & 10.90 &\textbf{ 0.13} & 58.7 & 85.5 & 0.67 & 12.27 &\textbf{ 0.13} & 29.9 & 61.7 & 0.40 & 71.37 & 1.00 & 41.0 & 66.1 & 0.49 & 79.52 & 1.00 \\
RREA       & 63.3 & 91.4 & 0.73 & 136.32 & 4.07  & 75.5 & 94.9 & 0.82 & 156.85 & 4.07  & --    &  --   &  --   &--  & -- &      --  &   --  &  --   & --  & -- \\
Dual-AMN       & 64.6 & \textbf{91.5} & 0.74 & 12.50 & 4.05  & 76.5 & \textbf{95.2} & 0.83 & 13.72 & 4.05  & 49.3    &  77.5  &  0.59   &413.73  & 21.91 &  59.3  &  81.8  &  0.67   & 456.87  & 22.56 \\
\hline
LargeEA-G & 30.0 & 63.5 & 0.39 &\textbf{ 9.76}  & \textbf{0.13} & 40.3 & 68.7 & 0.50 & \textbf{9.94} & \textbf{0.13} & 19.5  & 45.7 & 0.28 &\textbf{ 36.65 }&  \textbf{0.50}  & 21.4 & 39.7 & 0.27 & \textbf{39.29} & \textbf{0.50} \\
LargeEA-R & 47.2 & 74.0 & 0.56 & 41.01  & 1.01 & 58.0 & 77.6 & 0.65 & 42.12  & 1.01  & 32.5 & 55.5 & 0.40 & 163.92  & 4.04 & 27.3 & 42.5 & 0.32 & 157.15 & 4.04 \\
LargeEA-D & 45.7 & 69.6 & 0.54 & 13.51  & 0.75 & 58.7 & 76.2 & 0.65 & 13.96 & 0.75 & 33.0 & 54.8 & 0.40 & 134.5  & 3.41  & 28.6 & 42.7 & 0.65 & 121.9  & 3.65  \\
GCN-Align-S & 37.1 & 73.8 & 0.49 & 24.10  & 1.11 & 53.6 & 83.4 & 0.63 & 23.73 & 1.11 & 25.3 & 45.5 & 0.35 & 184.43  & 1.72  & 35.5 & 61.5 & 0.44 & 189.12  & 1.72  \\
RREA-S & 62.7 & 90.3 & 0.72 & 34.09  & 4.89 & 76.3 & 95.0 & 0.83 & 34.23 & 5.01 & 46.4 & 75.4 & 0.56 & 250.80  & 7.16  & 57.3 & 80.6 & 0.65 & 256.35  & 8.50 \\
Dual-AMN-S & 60.9 & 88.9 & 0.71 & 15.59  & 5.30 & 75.0 & 94.3 & 0.82 & 15.64  & 5.10 & 48.2 & 76.6 & 0.57 & 122.78  & 7.79  & 58.8 & 81.4 & 0.66 & 124.49  & 8.13 \\
\textbf{\ClusterEA{-G}} & 46.6 & 77.8 & 0.57 & 40.99  & 4.43 & 62.0 & 86.3 & 0.70 & 40.47 & 4.43 & 30.6 & 57.9 & 0.40 & 236.37 & 2.77  & 41.4 & 64.3 & 0.49 & 246.90 & 2.77  \\
\textbf{\ClusterEA{-R}} & \textbf{67.9} & 90.0 & \textbf{0.76} & 52.29  & 4.89 & 79.4 & 94.5 & \textbf{0.85} & 51.89  & 5.01  & 52.0 & 76.3 & 0.60 & 329.78  & 7.16 & 62.2 & 81.7 & 0.69 & 339.54 & 8.50 \\
\textbf{\ClusterEA{-D}} & 67.4 & 89.5 & 0.75 & 35.76 & 5.30 & \textbf{79.5} & 94.6 & \textbf{0.85} & 35.82  & 5.10  & \textbf{54.2} & \textbf{78.1} & \textbf{0.62} & 210.50  & 8.52 & \textbf{63.7} & \textbf{82.8} & \textbf{0.70} & 212.80 & 8.31 \\
\bottomrule
\end{tabular}}
\begin{tablenotes}
\footnotesize
    \item[1] The symbol ``--'' indicates that the model fails to perform EA on IDS100K dataset due to extensive GPU memory usage.
\end{tablenotes}
\end{threeparttable}
\vspace{-4mm}
\end{table*}

\noindent
\textbf{Datasets.}
We conduct experiments on datasets with different sizes from two cross-lingual EA benchmarks, i.e., IDS~\cite{OpenEA2020VLDB} and DBP1M~\cite{LargeEA22}.
\begin{itemize}[topsep=0pt,itemsep=0pt,parsep=0pt,partopsep=0pt,leftmargin=*]
\item
\emph{IDS} contains four cross-lingual datasets, i.e., English and French (IDS15K$_{EN-FR}$ and IDS100K$_{EN-FR}$), and English and German (IDS15K$_{EN-DE}$ and IDS100K$_{EN-DE}$). \MARK{These benchmarks are sampled with consideration of keeping the properties (e.g., degree distribution) consistent with their source KGs.} We use the latest 2.0 version of IDS, where the URIs of entities are encoded to avoid possible name bias.
\item \emph{DBP1M} is the largest cross-lingual EA benchmark. It contains two large-scale datasets extracted from DBpedia~\cite{DBPedia}, i.e., English and French (DBP1M$_{EN-FR}$), and English and German (DBP1M$_{EN-DE}$).
However, DBP1M is biased with name information. Specifically, part of the entities in inter-language links (ILLs) does not occur in the two KGs. Thus, we remove those ILLs to solve the name bias issue while retaining all the triples.
\end{itemize}

Following previous studies, we use 30\% of each dataset as seed alignment, and use 70\% of it to test the EA performance. As can be seen, we consider both degree distribution issue~\cite{RSN19, OpenEA2020VLDB} and the name bias issue~\cite{AttrGNN20} when selecting benchmarks, which meets the requirements of real-world applications.
Table \ref{tb:dataset} in Appendix~\ref{app:dataset-stat} lists the detailed information of the datasets used in our experiments.

\noindent
\textbf{Evaluation metrics.}
We use the widely-adopted \HitNFull{} (\HitN{}) and Mean Reciprocal Rank (\MRR{}) to verify the accuracy of \ClusterEA{}~\cite{MTransE17,IPTransE17, GCN-Align18, KECG19, RREA20, DualAMN21, EASY21}. Here, for \HitN{}, $N$=1, 10.
Higher \HitN{} and \MRR{} indicate better performance.
Also, we use running time and maximum GPU Memory cost (\emph{Mem.}) to evaluate the scalability of \ClusterEA{}. Specifically, running time is measured in seconds, and \emph{Mem.} is measured in Gigabytes.

\noindent
\textbf{Baselines.}
\MARK{
We compare \ClusterEA{} with structure-only based methods. If a baseline includes side information components, we remove them in order to guarantee a fair comparison
~\cite{DualAMN21, EVA20, RREA20, HyperKA20, OpenEA2020VLDB, LargeEA22}.  All the baselines are enhanced with CSLS (\SparseCSLS{} for large-scale datasets) before evaluation if possible. The implementation details and parameter settings of \ClusterEA{} and all baselines are presented in Appendix~\ref{app:details}.
Considering the scalability of models, we divide the compared baselines into two major categories, as listed below:
\begin{itemize}[topsep=0pt,itemsep=0pt,parsep=0pt,partopsep=0pt,leftmargin=*]
    \item \emph{Non-scalable baselines} that includes
            (i) \emph{GCNAlign}~\cite{GCN-Align18}, the first GNN-based EA model;
            (ii) \emph{RREA}~\cite{RREA20}, a GNN-based EA model that utilizes relational reflection transformation to obtain relation-specific embeddings for each entity, and is used as the default of LargeEA~\cite{LargeEA22}; and
            (iii) \emph{Dual-AMN}~\cite{DualAMN21}, a SOTA EA model that contains Simplified Relational Attention Layer and Proxy Matching Attention Layer for modeling both intra-graph and cross-graph relations.
    \item \emph{Scalable baselines} that includes
            (i) \emph{LargeEA}~\cite{LargeEA22}, the first EA framework that focuses on scalability by training multiple EA models on mini-batches generated by a rule-based strategy, and excepting for the two variants presented in~\cite{LargeEA22},
            we provide a new variant \emph{LargeEA-D} that incorporates recently proposed EA model Dual-AMN; and
            (ii) \emph{Stochastic training (Section~\ref{sec:mini-batch-training}) variant of GNN models}, which incorporates EA models with Stochastic training, including \emph{GCNAlign-S} for GCNAlign~\cite{GCN-Align18}, \emph{RREA-S} for RREA~\cite{RREA20}, and \emph{Dual-AMN-S} for Dual-AMN~\cite{DualAMN21}.
\end{itemize}
}

\noindent
\textbf{Variants of \ClusterEA{}.}
Since \ClusterEA{} is designed to be integrated with GNN-based EA models, we present three versions of \ClusterEA{}, i.e.,  \ClusterEA{-G} that includes GCNAlign, \ClusterEA{-R} that incorporates RREA, and  \ClusterEA{-D} that incorporates Dual-AMN. Specifically, we treat \ClusterEA{-D} as the default setting.

\vspace*{-3mm}
\subsection{Overall Results}

\begin{table}[t]\small
\caption{Overall EA results on DBP1M}\label{exp:overall_1M}
\vspace{-4mm}
\setlength{\tabcolsep}{0.1mm}{
\begin{tabular}{l|ccccc|ccccc}
\toprule
\multicolumn{1}{c|}{\multirow{2}{*}{Methods}} & \multicolumn{5}{c|}{DBP1M$_{EN-FR}$}  & \multicolumn{5}{c}{DBP1M$_{EN-DE}$}  \\ \cline{2-11}
\multicolumn{1}{c|}{} & \multicolumn{1}{c}{H@1} & \multicolumn{1}{c}{H@10} & \multicolumn{1}{c}{MRR} & \multicolumn{1}{c}{Time} &  \multicolumn{1}{c|}{Mem.} & \multicolumn{1}{c}{H@1} & \multicolumn{1}{c}{H@10} & \multicolumn{1}{c}{MRR} & \multicolumn{1}{c}{Time} & Mem. \\ \hline
LargeEA-G & 5.1 & 13.4 & 0.08 & \textbf{463}  & \textbf{4.00}   & 3.4 & 9.5 & 0.05 & \textbf{378} & \textbf{4.00}\\
LargeEA-R & 9.4 & 21.5 & 0.13 & 1681  & 20.05  & 6.4 & 15.0 & 0.09 & 1309 & 20.49 \\
LargeEA-D & 10.5 & 21.9 & 0.15 & 2546  & 19.72  & 6.6 & 14.7 & 0.09 & 1692 & 17.71 \\
GCN-Align-S & 6.0 & 19.1  & 0.10 & 1817  & 9.25 & 4.5 & 14.4  & 0.08 & 1491  & 7.78 \\
RREA-S & 21.1 & 42.2  & 0.28 & 2080  & 16.86  & 20.5 & 40.2  & 0.27 & 1754  & 15.65 \\
Dual-AMN-S & 22.4 & 43.4  & 0.29 & 588  & 15.85  & 22.0 & 41.7  & 0.28 & 490  & 14.61\\
\textbf{\ClusterEA{-G}} & 10.0 & 24.5 & 0.15 & 2501  & 17.43  & 6.9 & 17.7 & 0.11 & 2027  & 17.76 \\
\textbf{\ClusterEA{-R}}& 26.0 & 45.6  & 0.32 & 3025  & 21.11& 25.0 & 45.0  & 0.32 & 2526  & 19.68 \\
\textbf{\ClusterEA{-D}}& \textbf{28.1} & \textbf{47.4}  & \textbf{0.35} & 1647  & 20.10  & \textbf{28.8} & \textbf{48.8}  & \textbf{0.35} & 1360 & 18.20 \\
\bottomrule
\end{tabular}}
\vspace*{-6.0mm}
\end{table}

\noindent
\textbf{Performance on IDS.}\label{sec:exp_ids}
Table~\ref{exp:overall_15K_100K} summarizes the EA performance on IDS15K and IDS100K.
First,  \ClusterEA{} improves \HitOne{} by   $3.3\%-29.7\%$ compared with the non-scalable methods (viz., GCNAlign, RREA, and Dual-AMN), and by $3.2\%-42.3\%$ compared against the scalable ones. They validate the accuracy of \ClusterEA{}.
%
Moreover, all variants of \ClusterEA{} perform better than the structure-only based models in terms of \HitOne{}. It confirms the superiority of the way how we enhance the output, compared with CSLS that is the default setting for most recent EA models~\cite{EVA20,EASY21, MRAEA20,RREA20, OpenEA2020VLDB, DualAMN21} to enhance the output similarity matrix.
Second, it is observed that the accuracy of LargeEA variants is significantly reduced compared with their corresponding original models. This is because LargeEA discards structure information during its mini-batch training process. Unlike the LargeEA variants, the accuracy of Stochastic Training variants only drops slightly compared to the original non-scalable models. This shows that the Stochastic training process can minimize the structure information loss (cf. Section~\ref{sec:mini-batch-training}).
Third, as observed,  \HitTen{} of some non-scalable models (RREA, Dual-AMN) is higher than that of \ClusterEA{}. This is mainly due to (i) the information loss during \ClusterEA{}'s Stochastic training and (ii) the incompleteness of global similarity normalization (to be detailed in Section~\ref{sec:ablation}). Nevertheless, \HitOne{} is the most representative indicator for evaluating the accuracy of EA since higher \HitOne{} directly indicates more correct proportion of aligned entities. Thus, achieving the highest \HitOne{} among all the competitors is sufficiently to demonstrate the high performance of \ClusterEA{}.
Finally, the experimental results show that the training of all LargeEA variants is faster than that of other variants of the corresponding models. The reason is that LargeEA omits most information of graph structure, where less data is subjected to the training process. However, in this case, the running time of \ClusterEA{} is still comparable.

\noindent
\textbf{Performance on DBP1M.}
Table~\ref{exp:overall_1M} reports the EA performance of \ClusterEA{} and its competitors on DBP1M. Note that we do not report the results of the non-scalable models because it is infeasible to perform their training phases on DBP1M due to large GPU memory usage.
We observe that the accuracy of all the variants of \ClusterEA{} on DBP1M  is higher than those of LargeEA. Specifically, \HitOne{} is improved by $5.7\%-23.0\%$ and $6.8\%-25.4\%$ on DBP1M$_{EN-FR}$ and DBP1M$_{EN-DE}$, respectively.
Next, \MARK{compared with each Stochastic training variants of the corresponding incorporated model of \ClusterEA{}, \ClusterEA{} brings about $4.0\%-5.7\%$ and $2.4\%-6.8\%$ absolute improvement in \HitOne{} on DBP1M$_{EN-FR}$ and DBP1M$_{EN-DE}$ datasets, respectively. As the expressiveness of the model improves, \ClusterEA{} also offers more improvement.
}
Last, all the scalable variants incorporating GCNAlign variants produce poor EA results. Specifically, GCNAlign's model cannot be directly applied to large-scale datasets due to its insufficient expressive ability. \MARK{Note that \ClusterEA{} not only outperforms baselines in terms of accuracy but also achieves comparable performance with them in terms of both Mem. and running time.}
\MARK{Overall, \ClusterEA{} is able to scale-up the GNN-based EA models while enhancing \HitOne{} by at most $8\times$
compared with the state-of-the-arts. More experimental results of scalability are provided in  Appendix~\ref{app:scalability}.}

\vspace*{-3mm}
\begin{figure*}[t]
\centering
\includegraphics[width=.8\textwidth]{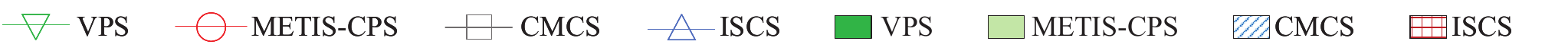}\\
\vspace*{-4mm}
\subfigure[$\rm IDS15K_{EN-FR}$]{
 \includegraphics[width=1.7in]{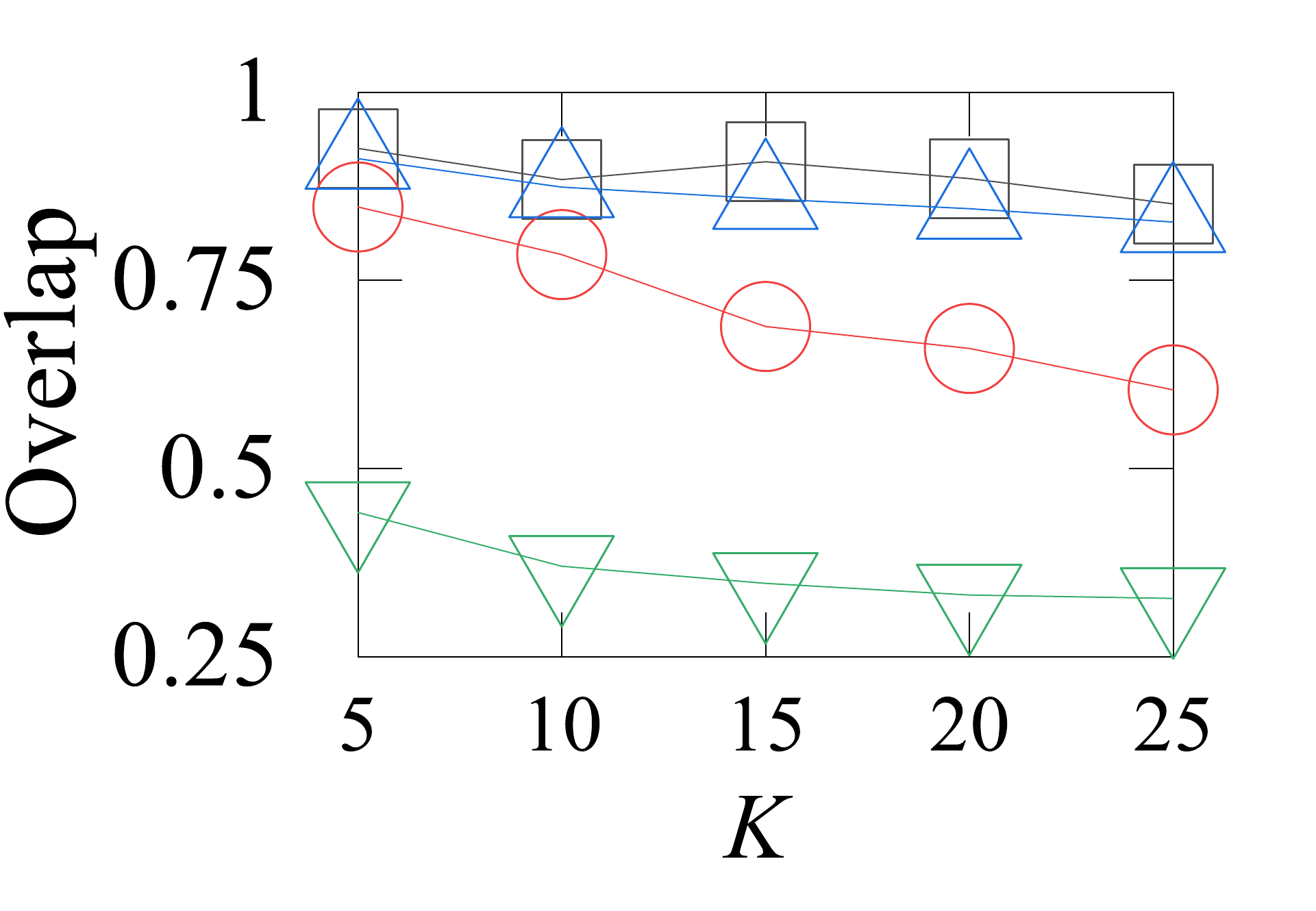}
 \label{fig:partition-k-small-fr}
}\hspace{-3mm}
\subfigure[$\rm IDS100K_{EN-FR}$]{
 \includegraphics[width=1.7in]{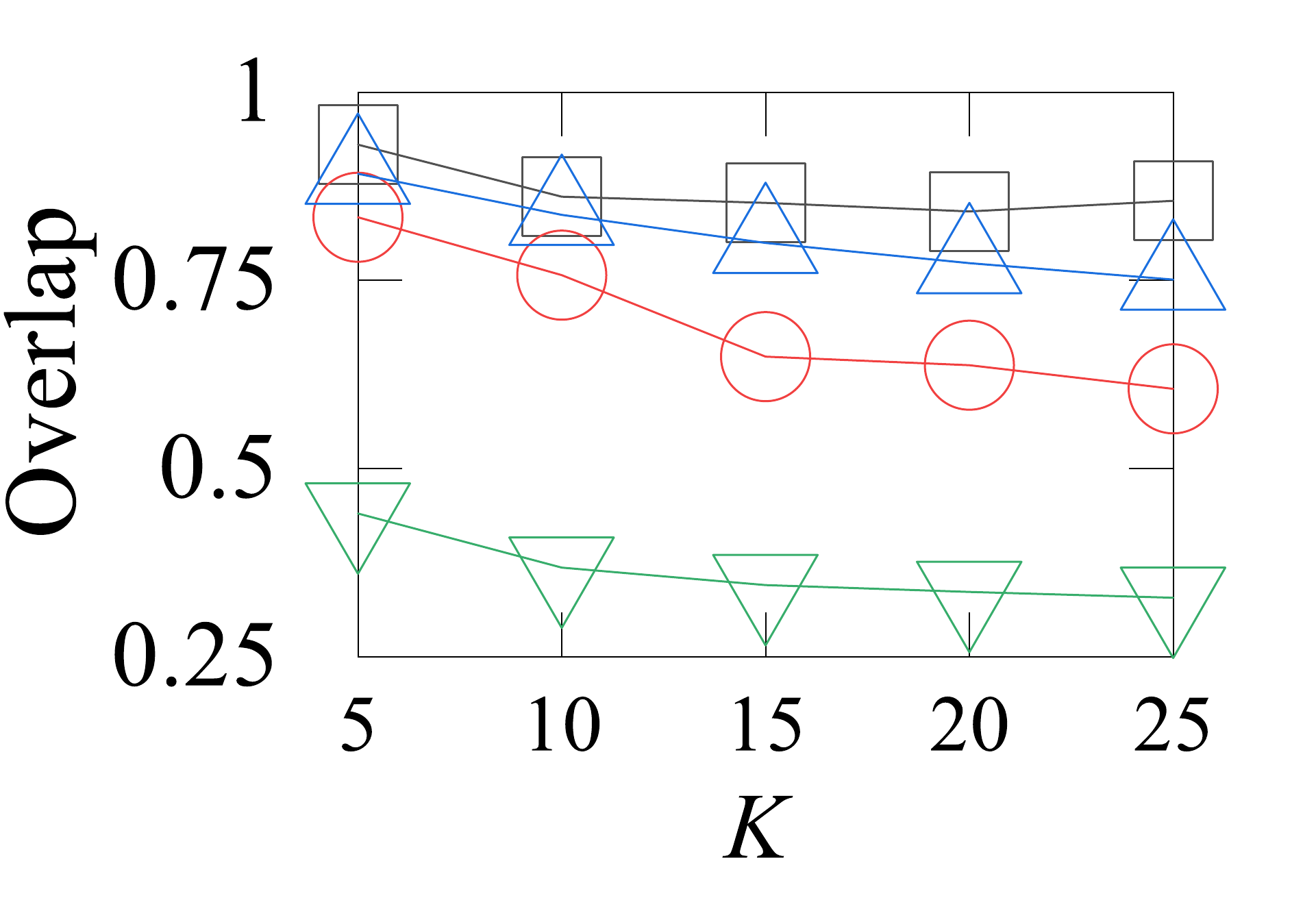}
 \label{fig:partition-k-medium-fr}
}\hspace{-3mm}
\subfigure[$\rm DBP1M_{EN-FR}$]{
 \includegraphics[width=1.7in]{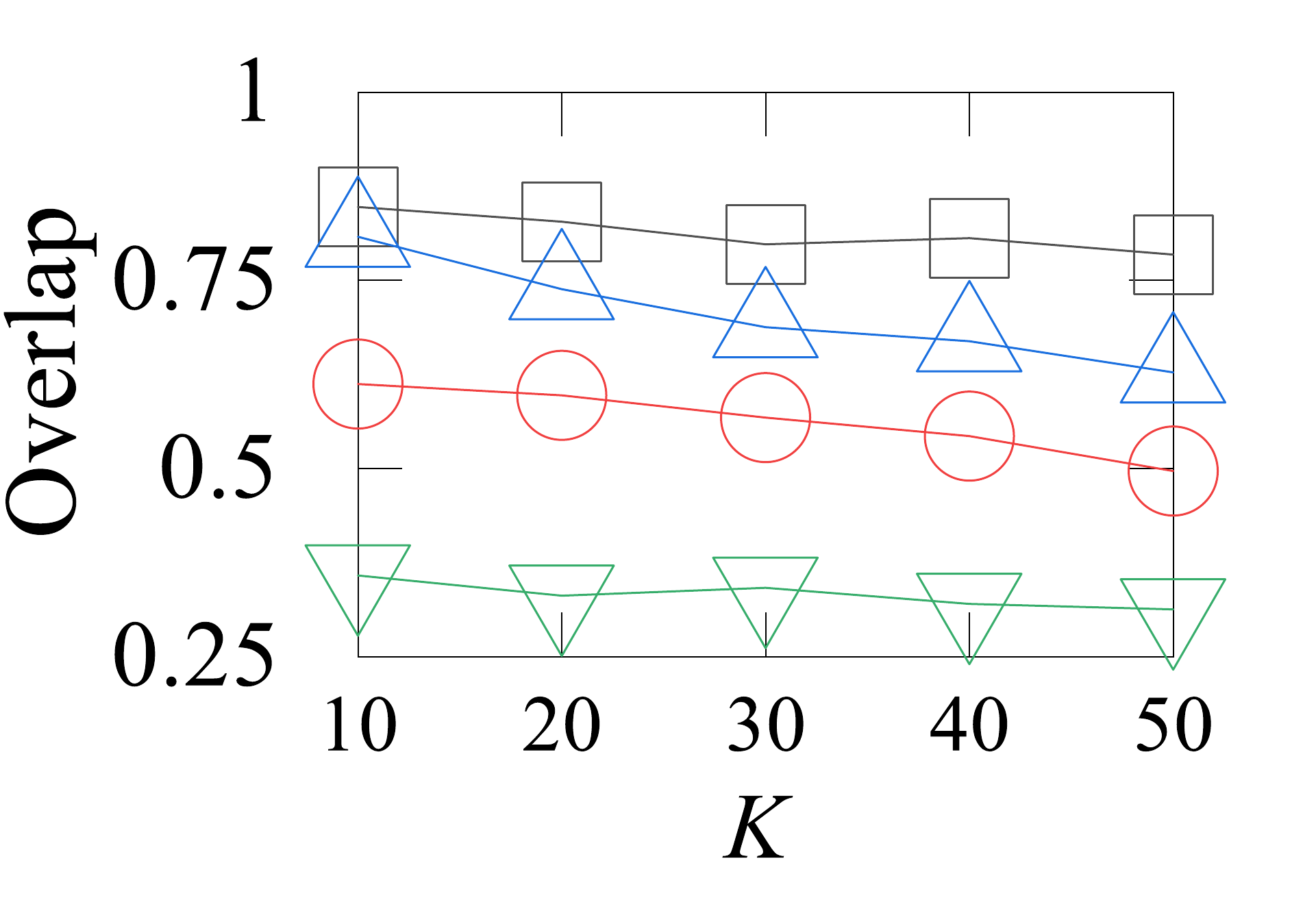}
 \label{fig:partition-k-large-fr}
}\hspace{-3mm}
\subfigure[$\rm DBP1M_{EN-FR}$]{
 \includegraphics[width=1.67in]{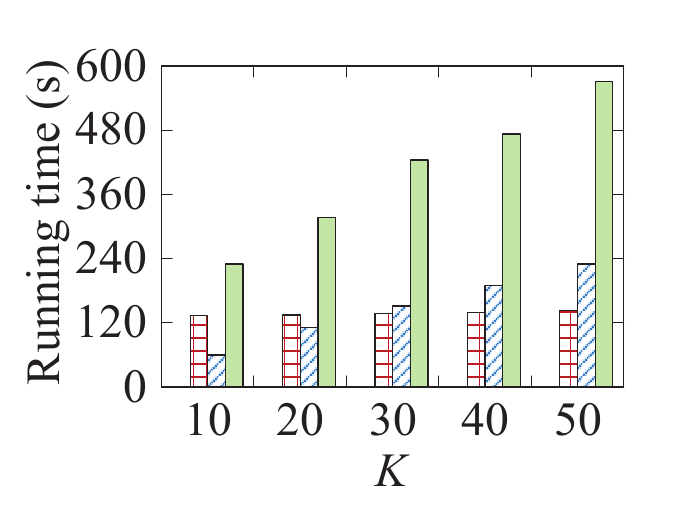}
 \label{fig:scalability-p-fr}
}\\
\vspace*{-6mm}
\subfigure[$\rm IDS15K_{EN-DE}$]{
 \includegraphics[width=1.7in]{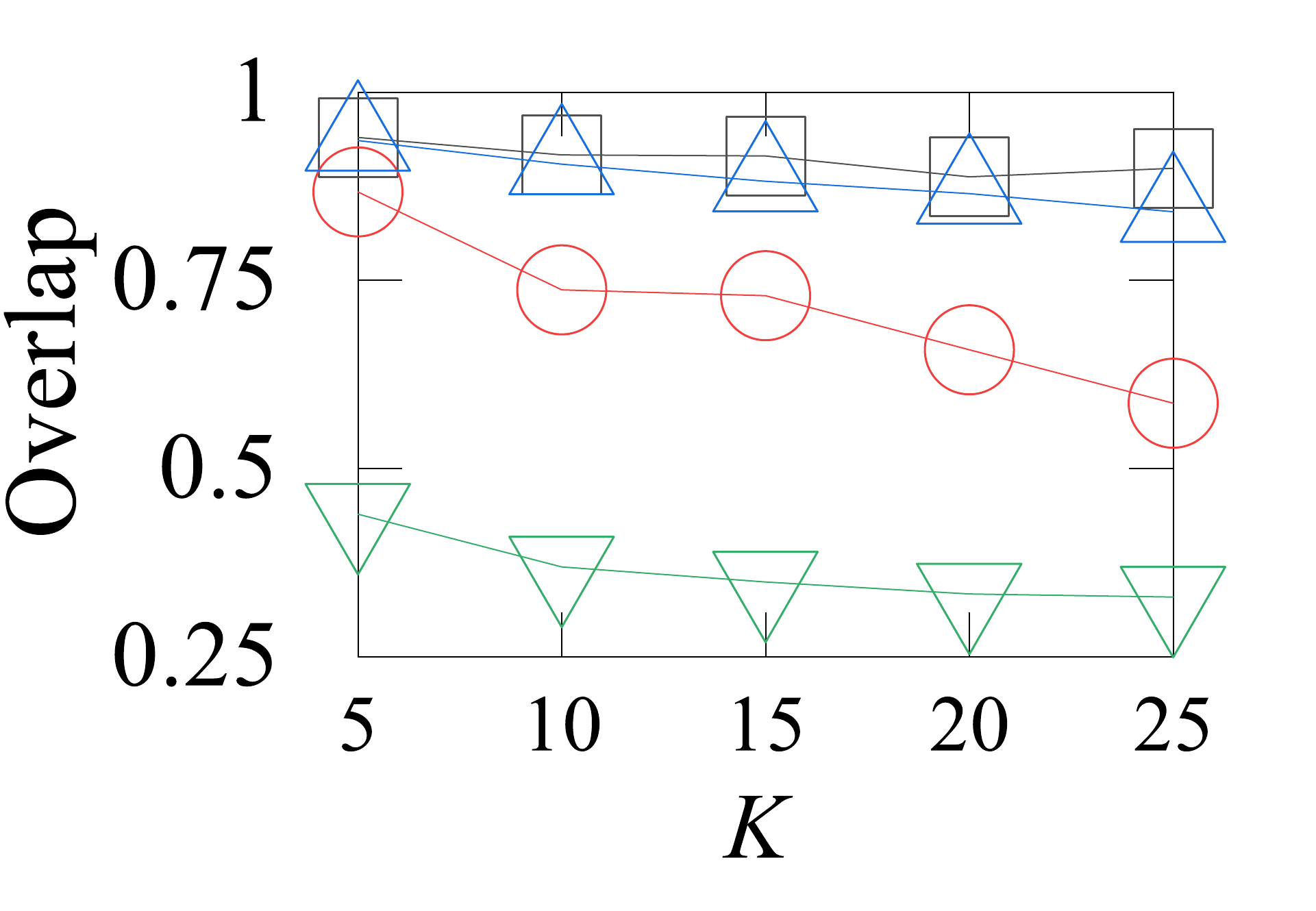}
 \label{fig:partition-k-small-de}
}\hspace{-3mm}
\subfigure[$\rm IDS100K_{EN-DE}$]{
 \includegraphics[width=1.7in]{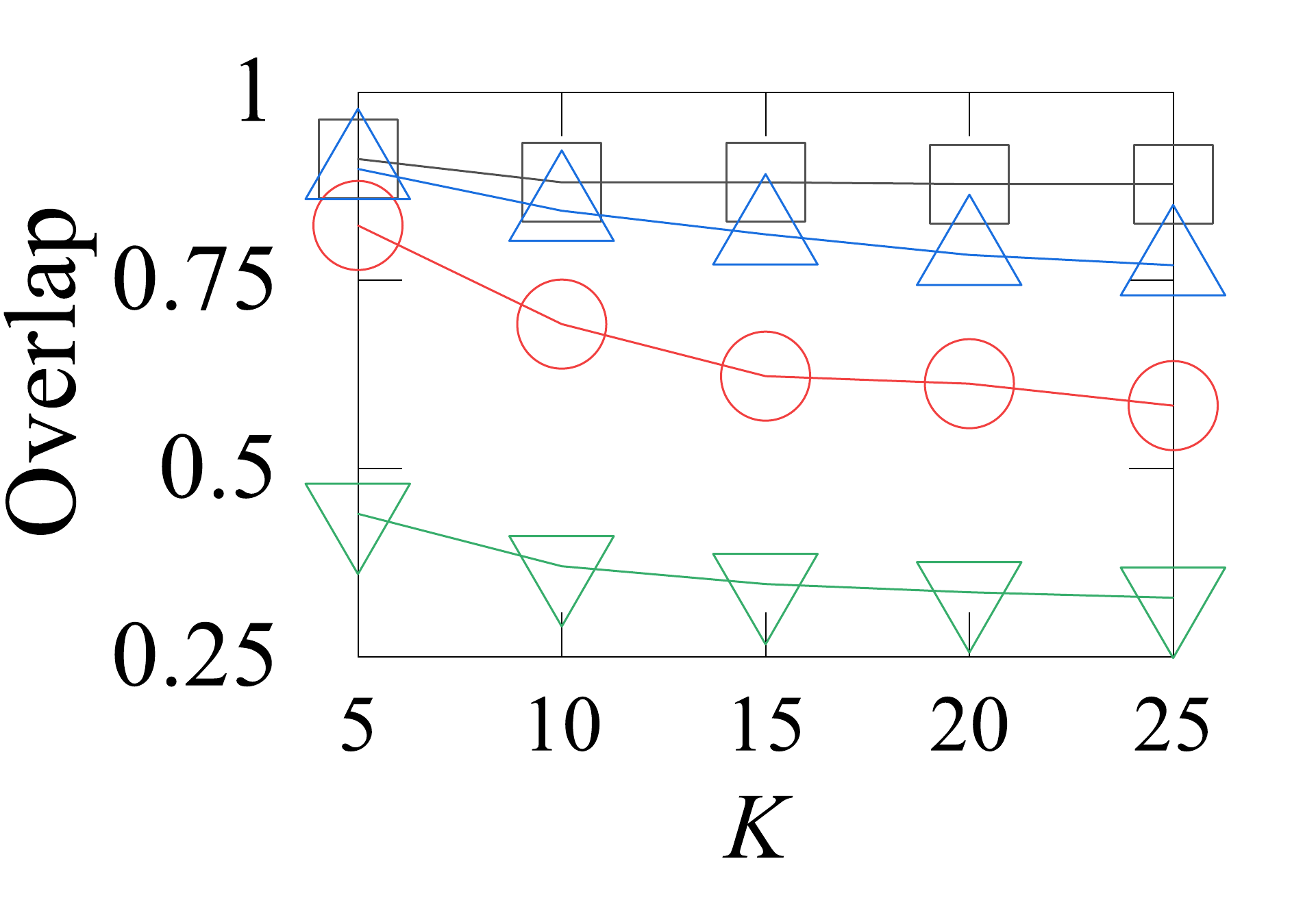}
 \label{fig:partition-k-medium-de}
}\hspace{-3mm}
\subfigure[$\rm DBP1M_{EN-DE}$]{
 \includegraphics[width=1.7in]{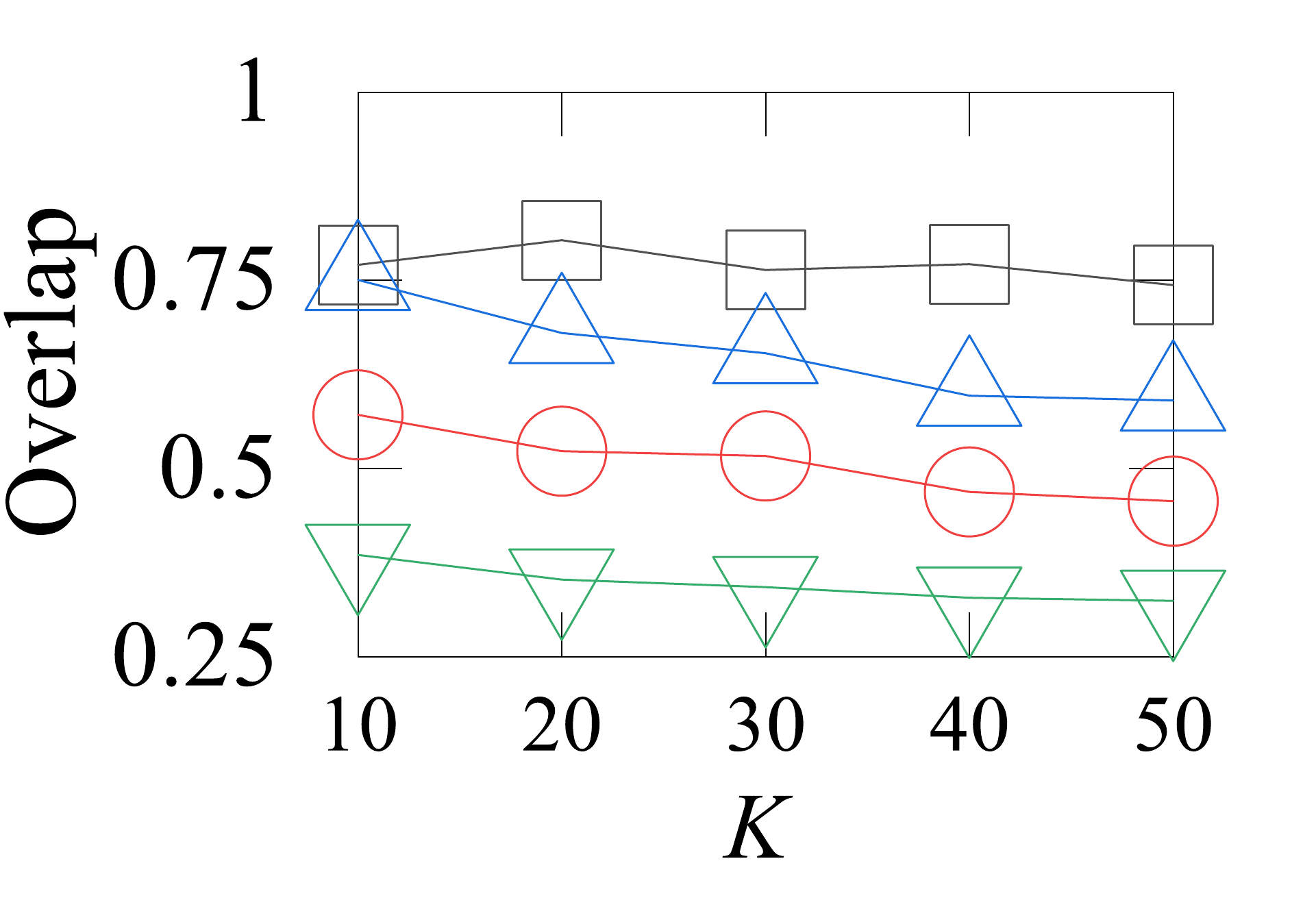}
 \label{fig:partition-k-large-de}
}\hspace{-3mm}
\subfigure[$\rm DBP1M_{EN-DE}$]{
 \includegraphics[width=1.67in]{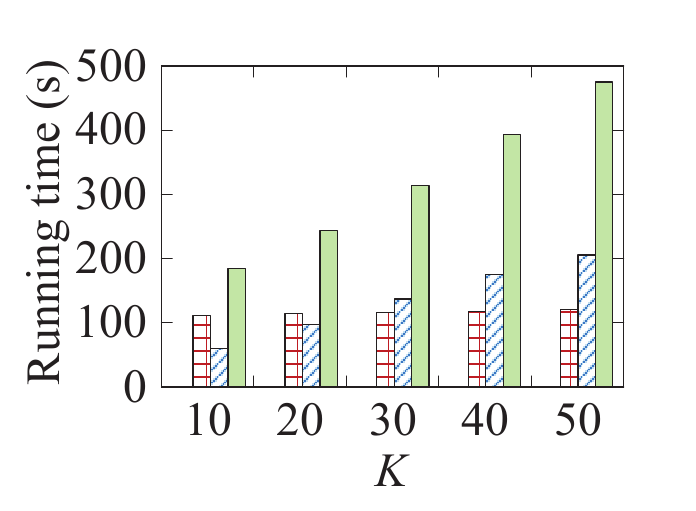}
 \label{fig:scalability-p-de}
}
\vspace{-4mm}
\caption{Comparison results of different batch sampling strategies}
\vspace*{-4mm}
\label{fig:partition-exp}
\end{figure*}

\begin{table}[t]\small
\caption{The result of ablation study}\label{exp:abl}
\vspace*{-4mm}
\setlength{\tabcolsep}{1mm}{
\begin{tabular}{l|ccc|ccc}
\toprule
\multicolumn{1}{c|}{\multirow{2}{*}{Methods}} & \multicolumn{3}{c|}{DBP1M$_{EN-FR}$}  & \multicolumn{3}{c}{DBP1M$_{EN-DE}$}  \\ \cline{2-7}
\multicolumn{1}{c|}{} & \multicolumn{1}{c}{H@1} & \multicolumn{1}{c}{H@10} & \multicolumn{1}{c|}{MRR} & \multicolumn{1}{c}{H@1} & \multicolumn{1}{c}{H@10} & \multicolumn{1}{c}{MRR}  \\ \hline
\ClusterEA{}                    & \textbf{28.1} & 47.4 & \textbf{0.35} & \textbf{28.8} & \textbf{48.8} &  \textbf{0.35}\\
\ClusterEA{} - \SparseCSLS{}    & 27.9 & 46.2 & 0.34 & 28.4 & 46.0 &  0.34\\
\ClusterEA{} - Global Sim       & 27.6 & \textbf{48.3} & 0.34 & 28.3 & 48.7 &  \textbf{0.35}\\
\ClusterEA{} - Sinkhorn         & 20.0 & 46.1 & 0.29 & 22.7 & 46.2 &  0.30\\
\ClusterEA{} - \KMeans{}        & 26.5 & 46.2 & 0.33 & 26.8 & 47.1 &  0.33\\
\ClusterEA{} - \MetisGCN{}      & 25.3 & 45.9 & 0.32 & 25.5 & 46.5 &  0.32\\
\ClusterEA{} - Dual-AMN         & 10.0 & 24.5 & 0.15 & 6.9  & 17.7  &  0.11\\
\bottomrule
\end{tabular}}
\vspace*{-4mm}
\end{table}

\subsection{Ablation Study}\label{sec:ablation}
In ablatoin study, we remove
each component of \ClusterEA{}, and report \HitOne{}, \HitTen{}, and \MRR{} in Table~\ref{exp:abl}.
First, after removing the \SparseCSLS{} component, the accuracy of \ClusterEA{} drops. This shows that the \SparseCSLS{} normalization indeed address the geometric problems of global similarity on large-scale EA.
Second, after removing the global similarity, \HitOne{} of \ClusterEA{} drops but \HitTen{} grows on DBP1M$_{EN-FR}$. The fluctuation on \HitTen{} may be due to the incompleteness of global similarity normalization
that disturbs the final similarity matrix.
Specifically,
the local similarity is normalized into nearly permutation matrices with Sinkhorn iteration, where most values of one row are close to zero.
When fusing local and global similarity matrices, elements that are not top-1 in the local matrix will be biased towards the value of the global matrix, which is only partially normalized.
This causes the disturbance.
However,
the incompleteness of normalization does not degrade \HitOne{}. This is because the value of one row in $\mathcal{M}_{L}$ that is correctly aligned will be normalized into a higher value, providing resistance to the global matrix disturbance.
Third, after removing the Sinkhorn iteration,  the accuracy of \ClusterEA{} drops significantly. This confirms the importance of normalizing the similarity matrices. Finally, after removing each sampler of \Sampling{}, the accuracy of \ClusterEA{} drops on all metrics. This validates the importance of fusing information from multi-aspects, including cross-KG information and intra-KG information.
In addition, we observe that \KMeans{} has less influence compared with \MetisGCN{}. This is mainly due to the following two reasons. First, \KMeans{} generally clusters entities with similar embedding vectors, where each batch still suffers from geometric problems. On the contrary,  \MetisGCN{} samples batches based on the graph neighborhood information, which is a strong constriction on the learning model. Second, we apply \MetisGCN{} in both directions. In this case, it can capture more information than \KMeans{}.
By replacing its Dual-AMN model with the GCNAlign model, the accuracy of \ClusterEA{} drops significantly on all metrics. This demonstrates the importance of the ability of the incorporated EA model in \ClusterEA{}.

\vspace{-4mm}
\subsection{Case Study: \Sampling{} Analysis}
\label{sec:exp_minibatchgeneration}

The \Sampling{} is a vital component of \ClusterEA{}. To ensure scalability, we set the batch number $K$ such that the space cost of the normalization process does not exceed the GPU memory. We also need to guarantee that our batch sampling method can produce acceptable mini-batches under different $K$ settings.
Thus, we provide a detailed analysis on varying the batch number $K$ for different batch samplers from \Sampling{} (i.e., \MetisGCN{} and \KMeans{}). Specifically, we study how much are the mini-batches generated by one sampler acceptable as the percentage of equivalent entities that are placed into the same mini-batches (denoted as \textit{Overlap}).
Next, we report the Overlap metric and running time of the proposed sampler, where the mini-batch number $K$ of the proposed sampler is varied from 5 to 25 on IDS and is varied from 10 to 50 on DBP1M. We compare the proposed sampler with two rule-based baselines, VPS and METIS-CPS. \emph{VPS} randomly partitions seed alignments and all other entities into different mini-batches. \emph{METIS-CPS} sets the training entities with higher nodes to sample better mini-batches.
Note that both METIS-CPS and \MetisGCN{} are unidirectional. Thus, we apply these methods in both directions, and present their average performance of the two directions. We report the overlap of all sampling methods on all benchmarks in Fig~\ref{fig:partition-exp}, where Figures~\ref{fig:partition-k-small-fr},~\ref{fig:partition-k-medium-fr},~\ref{fig:partition-k-large-fr},~\ref{fig:partition-k-small-de}, ~\ref{fig:partition-k-medium-de}, and~\ref{fig:partition-k-large-de} are the overlap of different datasets. The results show that \KMeans{} generally outperforms two baselines on all the datasets, and its performance is stable when varying $K$. However, although it is better overlapped, it may incur hubness and isolation problems in mini-batches (cf. Section~\ref{sec:ablation}). Thus, fusing multi-aspect is essential for \ClusterEA{}. \MetisGCN{} results in less overlapped mini-batches while still much better than METIS-CPS. This is because METIS does not necessarily follow the guidance provided by METIS-CPS. The GNN node classification model in \MetisGCN{}, which considers the cross-entropy loss as a penalty, is forced to learn mini-batches more effectively. Since we set the training ratio to 30\%, all samplers have an overlap over 30\%, including VPS that splits mini-batches randomly.
Finally, we report the running time on DBP1M in Figures~\ref{fig:scalability-p-fr} and~\ref{fig:scalability-p-de}. Since all samplers achieve sufficiently high efficiency on IDS datasets, we do not present the running time of all samplers on IDS due to space limitation. It is observed that, although the result is unacceptable, VPS is the fastest sampling method. Both the proposed \MetisGCN{} and \KMeans{} are always about $2\times$ faster than the rule-based METIS-CPS when changing $K$. The reason is that both of them utilize machine learning models that can be easily accelerated with GPU.

\vspace{-2mm}
\section{Conclusions}
\label{sec:conclusions}

We present \ClusterEA{} to align entities between large-scale knowledge graphs with stochastic training and normalized similarities.
\ClusterEA{} contains three components, including stochastic training, \Sampling{}, and \Merging{}, to perform large-scale EA task based solely on structure information.
We first train a large-scale Siamese GNN for EA in a stochastic fashion to produce entity embeddings. Next, we propose a new \Sampling{} strategy for sampling highly overlapped mini-batches taking advantage of the trained embeddings. Finally, we present a \Merging{} process, which first normalizes local and global similarities and then fuses them to obtain the final similarity matrix. The whole process of \ClusterEA{} guarantees both high accuracy and comparable scalability.
Considerable experimental results on EA benchmarks with different scales demonstrate that \ClusterEA{} significantly outperforms previous large-scale EA study. In future, it is of interest to explore dangling settings~\cite{NoMatch21} of EA on large scale datasets. 

\vspace{-2mm}
\section{Acknowledgements}
This work was supported in part by the National Key Research and Development Program of China under Grant No. 2021YFC3300303, the NSFC under Grants No. (62025206, 61972338, and 62102351), and the Zhejiang Provincial Natural Science Foundation under Grant No. LR21F020005. Lu Chen is the corresponding author of the work.

\vspace{-2mm}
\bibliographystyle{ACM-Reference-Format}
\bibliography{REFER}
\pagebreak
\begin{appendices}

\section{Normalized Hard Sample Mining Loss}
\label{app:nhsm}
We detail the NHSM loss used in our work here.
 Formally, for the current mini-batch, the training loss is defined as
\begin{equation}
     L = LogSumExp(\lambda z(e_s, e_t)) + LogSumExp(\lambda z(e_t, e_s)) 
\end{equation} 
, where $e_s \in B_s$, $e_t \in B_t$, $(e_s, e_t) \in \phi^{\prime}$
, and
$LogSumExp(X) = log(\sum_{x \in X}e^x)$
is an operator to smoothly generate hard negative samples,
$\lambda$ is the smooth factor of $LogSumExp$, and 
$z \in\mathcal{R}^{|B_t|}$ is the normalized triple loss, defined as

\vspace{-5mm}
\begin{equation}
z(e_{s}, e_{t}) =\operatorname{z-score}(\{\gamma+\operatorname{sim}(e_{s}, e_{t})-\operatorname{sim}(e_{s}, e_{t}^{\prime}) | e^{\prime}_t \in B_t\}),
\end{equation}
where $\operatorname{z-score}(X)=  \frac{X - \mu(X)}{\sigma(X)}$ is the standard score normalization,
and $\operatorname{sim}(e_{s}, e_{t})= \mathbf{h}_{e_s} \cdot \mathbf{h}_{e_t}$ is the similarity of two entities obtained with the GNN output feature.

\section{Adjacency Matrix Construction for \MetisGCN{}}
\label{app:adj-matrix}
Following~\cite{GCN-Align18}, to construct the adjacency matrix $A \in R^{|E|\times|E|}$, we compute two metrics, which are called functionality and inverse functionality, for each relation:
\begin{equation}
\begin{aligned}
&\operatorname{fun}(r)=\frac{\#  Head\_Entities\_of\_r }{\# Triples\_of\_r } \\
&\operatorname{ ifun }(r)=\frac{\# Tail\_Entities\_of\_r }{\#  Triples\_of\_r }
\end{aligned}
\end{equation}
where $\#Triples\_of\_r$ is the number of triples of relation $r$, \\
$\#Head\_Entities\_of\_r$ is the number of head entities of $r$, and \\
$\#Tail\_Entities\_of\_r$ is the number of tail entities of $r$. To measure the influence of the $i$-th entity over the $j$-th entity, we set $a_{i j} \in A$ as:

\begin{equation}
a_{i j}=\sum_{\left\langle e_{i}, r, e_{j}\right\rangle \in G} \operatorname { ifun }(r)+\sum_{\left\langle e_{j}, r, e_{i}\right\rangle \in G}\operatorname{fun}(r)
\end{equation}

\section{Statistics of datasets}
\label{app:dataset-stat}

\begin{table}[t]
\centering
\caption{Statistics of IDS15K, IDS100K, and DBP1M}
\vspace{-4mm}
\label{tb:dataset}
\setlength{\tabcolsep}{0.4mm}{
\begin{tabular}{ll|l|l|l}
\toprule
\multicolumn{2}{c|}{Datasets}                          & \#Entities          & \#Relations & \#Triples           \\ \hline
\multicolumn{1}{l|}{\multirow{2}{*}{IDS15K}}  & EN-FR & 15,000-15,000       & 267-210     & 47,334-40,864       \\
\multicolumn{1}{l|}{}                         & EN-DE & 15,000-15,000       & 215-131     & 47,676-50,419       \\ \hline
\multicolumn{1}{l|}{\multirow{2}{*}{IDS100K}} & EN-FR & 100,000-100,000     & 400-300     & 309,607-258,285     \\
\multicolumn{1}{l|}{}                         & EN-DE & 100,000-100,000     & 381-196     & 335,359-336,240     \\ \hline
\multicolumn{1}{l|}{\multirow{2}{*}{DBP1M}}   & EN-FR & 1,877,793-1,365,118 & 603-380     & 7,031,172-2,997,457 \\
\multicolumn{1}{l|}{}                         & EN-DE & 1,625,999-1,112,970 & 597-241     & 6,213,639-1,994,876 \\ \bottomrule
\end{tabular}}
\end{table}

\begin{figure}[t]
\centering
\includegraphics[width=0.46\textwidth]{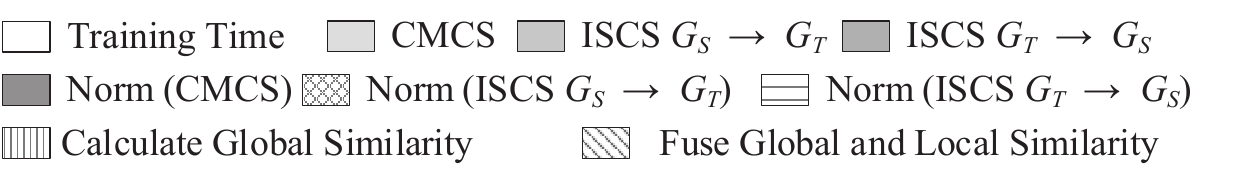}\vspace*{-4.5mm}\\
\subfigure[EN-FR]{
 \includegraphics[width=1.55in]{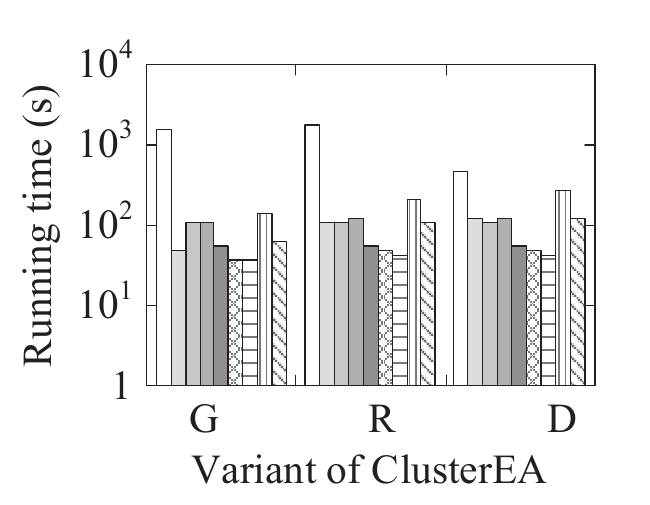}
}
\subfigure[EN-DE]{
 \includegraphics[width=1.55in]{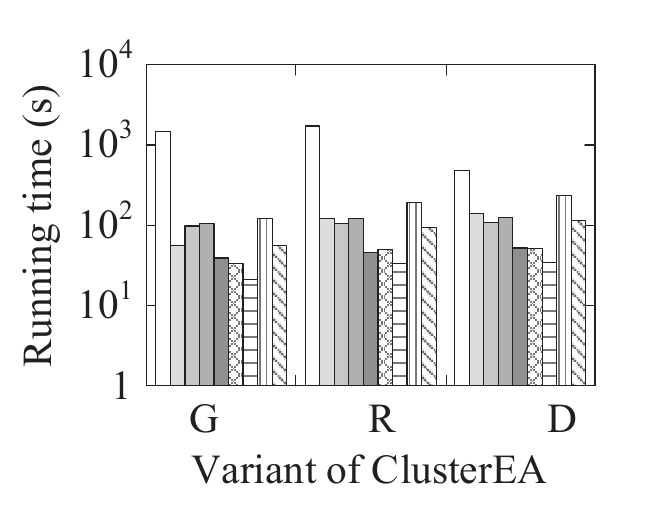}
}
\caption{Scalability analysis vs. variants of \ClusterEA{}}
\label{fig:scalability}
\end{figure}

\section{Implementation Details}
\label{app:details}
The results of all the baselines are obtained by our re-implementation with their publicly available source codes.
All experiments were conducted on a personal computer with an Intel Core i9-10900K CPU, an NVIDIA GeForce RTX3090 GPU, and 128GB memory. The programs were all implemented in Python.
Since the deep learning frameworks used vary on different implementations of EA models, we use the NVIDIA Nsight Systems\footnote{https://developer.nvidia.com/nsight-systems} to record the GPU memory usage of all approaches uniformly.
We detail the hyper-parameters used in the experiment as follows. All the hyper-parameters are set without special instructions.

\subsection{Non-scalable methods} For non-scalable methods, including GCNAlign, RREA, and Dual-AMN, we follow the parameter settings reported in their original papers~\cite{GCN-Align18,RREA20, DualAMN21}.

\subsubsection{GCNAlign}
Note that GCNAlign contains an attribute encoder as the side information component.
We re-implement the GCNAlign model by removing attribute features from it.

\subsection{LargeEA} We reproduce LargeEA by removing entirely the name channel. It is worth noting that LargeEA incorporates a name-based data augmentation process for generating alignment seeds other than training data. For fair comparison, we also remove this augmentation process, making the LargeEA framework not use any alignment signals apart from the training data. To be more specific, we only compare the structure-channel of LargeEA. The running time and maximum GPU memory usage are also recorded only for the structure-channel of LargeEA, including METIS-CPS partitioning and mini-batch training. We use the default settings reported in~\cite{LargeEA22} of the two versions, LargeEA-G and LargeEA-R. For the newly proposed variant LargeEA-D, we keep all the hyper-parameters unchanged except the structure-based EA model-related parameters switched to the default settings of Dual-AMN~\cite{DualAMN21}.

\subsection{Stochastic training variant of non-scalable methods} Contributed by the NHSM loss~\cite{DualAMN21}, the training epoch number could be greatly reduced. We replace the training epoch for GCNAlign-S and RREA-S from 2000 and 1200 to 50. We follow~\cite{DualAMN21} to train 20 epochs for Dual-AMN-S on IDS datasets. We further notice that the training loss of Dual-AMN-S is stable after 10 epochs for the large-scale dataset  DBP1M, thus setting the training epoch of Dual-AMN-S to 10 for DBP1M.
We set the fan out number in neighborhood sampling $F=8$, the batch size $N_p = 2000$, $N_n = 4000$, and Adam as the training optimizer for all variants. We follow the setup of~\cite{DualAMN21} for NHSM loss.
For other hyper-parameter settings of the model, including the embedding dimension $D$, we follow their original papers~\cite{GCN-Align18,RREA20, DualAMN21}.

\subsection{\ClusterEA{}}
In the \emph{stochastic training process}, we adopt the aforementioned settings for each \ClusterEA{} variant. In the \emph{\Sampling{} process}, we set the mini-batch number $K = 5$ for IDS15K, $K = 10$ for IDS100K, and $K = 30$ for DBP1M. Moreover, we utilize a cache in disk for pre-processed edge information to quickly load the constructed matrix of relation-aware GNN variants. For \KMeans{}, we set the max iteration number to $300$, the tolerance to $10^{-4}$, and distance metric to euclidean distance for the KMeans algorithm. We adopt the default setting for the XGBoost Classifier~\cite{XGBoost16}. We utilize GPU acceleration for those methods. For \MetisGCN{}, we adopt the default setting of METIS~\cite{METIS98}, and train a two layer GCN with Adam. The learning rate is set to $0.01$, and the training epoch of GCN set to $800$ for IDS15K, $1500$ for IDS100K, and $3000$ for DBP1M. In the \emph{\Merging{} process}, we set $K_s = 100$ for the iteration round of Sinkhorn, $K_r = 50$ for top-K similarity serach, and $K_n = 10$ for CSLS.

\section{Scalability Evaluation}
\label{app:scalability}

To further investigate the scalability of our proposed \ClusterEA{} framework, we verify the running time of each components in different variants of \ClusterEA{} on the DBP1M dataset, including (1) \emph{Training time} of the EA model; (2) \emph{\KMeans{}} batch sampling time; (3) \emph{\MetisGCN{}} batch sampling time on both directions; (4) \emph{Local Similarity Normalization time} of the batches generated by each samplers, denoted as Norm(Sampler); (5) \emph{Calculating Global Similarity} time in \Merging{}; and (6) \emph{Fusing Global and Local Similarity} time in \Merging{}. 
The experimental results are shown in Figure~\ref{fig:scalability}, where G, R, and D denote \ClusterEA{-G}, \ClusterEA{-R}, and \ClusterEA{-D}, respectively. As depicted in Figure~\ref{fig:scalability}, we observe that the running time of each component except for Training time does not exceed $10^3$ seconds on the large-scale dataset.
This confirms the scalability of \ClusterEA{}. Note that the training time of the Dual-AMN model is significantly less than that of other variants. This is contributed by the model design of Dual-AMN that could dramatically reduce the required training epochs.

\end{appendices}

\balance
\end{document}